\documentclass[smallextended]{svjour3}       
\smartqed  
\usepackage{amsmath}
\usepackage{graphicx}
\usepackage{caption}
\usepackage{subfigure}
\usepackage{float}

\begin{document}
\title{Consequences of tidal interaction between disks and orbiting protoplanets for the evolution of multi-planet systems
with architecture resembling that of Kepler 444}
 \subtitle{}

\titlerunning{ Disk-planet tidal interaction and  close orbiting  planetary systems }        

\author{
        J. C. B. Papaloizou }


\institute{ 
       J. C. B. Papaloizou  \at
DAMTP, Centre for Mathematical Sciences, University of Cambridge,
Wilberforce Road, Cambridge CB3 0WA, United Kingdom\\
\email{ J.C.B.Papaloizou@damtp.cam.ac.uk} }

\date{Received: date / Accepted: date}

\maketitle

\begin{abstract}
{We study  orbital evolution of multi-planet systems with masses in the terrestrial planet regime  induced through
tidal interaction with a protoplanetary disk assuming that this is the dominant mechanism for producing orbital migration and circularization.
We develop   a  simple analytic model for a  system   that maintains consecutive pairs in resonance while
undergoing orbital circularization and migration. 
 This model enables
migration times for each planet to be estimated  once planet masses, circularization times
and the migration time for the innermost planet are  specified. We applied it to a system with
the current architecture of Kepler 444 adopting a simple protoplanetary  disk model and planet masses that yield migration times
inversely proportional to the planet mass, as expected if they result from 
torques due to tidal interaction with the protoplanetary  disk.  Furthermore the  evolution time
for the system as a whole is  comparable to current protoplanetary disk lifetimes.

In addition we have performed  a number of numerical simulations with input data obtained from this model. These indicate that although the analytic
model is inexact, relatively small corrections to the estimated migration rates yield systems 
for which period ratios vary by a minimal extent.

Because of relatively large deviations from exact resonance  in the observed system  of up to $2\%,$  the migration times obtained in this way indicate 
 only  weak convergent migration such that a  system  for which the planets did  not interact would  contract  by only  $\sim 1\%$  although  
undergoing significant inward migration as a whole.
We have also performed additional simulations to investigate conditions under which the  system could undergo significant convergent migration before
reaching its  final state. These indicate that  migration times have to be significantly shorter and  resonances  between planet pairs significantly  closer
during such an evolutionary phase.
 Relative migration rates would then have to decrease allowing period  ratios to increase to become more distant from resonances
 as the system approached its final state in the inner regions of the protoplanetary disk. 
}

\end{abstract}



\section{Introduction}\label{sec1}

The Kepler mission has discovered an abundance of confirmed and candidate  planets 
orbiting close to their host stars ( Batalha et al. 2013).
Many of these are in highly compact  planetary systems.
A significant number  contain pairs that are  close to first order commensurabilities. 
 Lissauer et al. (2011)a  found  Kepler  candidates in short period orbits in  multi-resonant configurations.
For example  the  Kepler 223  system  contains  four planets   exhibiting   the mean motion ratios  8:6:4:3  and   
 Kepler 80  is a   near-resonant system  of five planets in which there are  two three body  mean motion resonances 
 $2n_2- 5n_3+3n_4\sim 0$ and     $2n_3-6n_4+4n_5\sim 0,$ with $n_i$ being the mean motion of planet $i.$  
In addition  the Kepler 60 system  has 
three planets with the inner pair very close to a 5:4 commensurability and the outer pair 
very close to a 4:3 commensurability  (Steffen et al. 2013) .

The recently confirmed  Kepler 444  system  contains   five transiting, sub-Earths (Campante et al. 2015).
This  system  is  highly  compact with
all planets  orbiting   the  parent  star  within  0.08 AU.
In addition the orbits  are consistent with  near-coplanarity.   Consecutive planet pairs are
close to first order orbital commensurabilities,  with relative deviations  of less than  
 $2 \%$  in excess of 
5:4,  4:3,  5:4,  and  5:4   resonances. 
The central  star is metal poor.  It and hence the planetary system has an age of 11:2  G y.,
 making it the oldest known system of terrestrial planets.
  Furthermore  it has a binary companion at a projected  separation of
 $\sim$ 60 AU (Campante et al. 2015).  If this  was  that close  when the protoplanetary disk was present,  it is likely to have  influenced planet formation and evolution.
 These features make it of special interest in the context of planet formation theories.

Tightly packed resonant planetary systems  of this kind are of interest  on account of what their dynamics can tell us
about their formation and  early orbital evolution.
The formation of resonant chains  in tightly packed  systems of planets
is expected to occur as a result  of convergent orbital 
migration produced by the action  of  torques resulting from tidal  interaction with the protoplanetary disk 
from  which they formed (eg. Cresswell \& Nelson 2006;  Terquem and Papaloizou 2007).

Migration due to tidal interaction with the disk has also been suggested as a  mechanism through which
planets end up on short period orbits in  order to avoid problems associated with in situ formation (  e.g.
Raymond et al. 2008). 
 However,  the total amount of  radial migration  does not need to be large  in order to  form  near commensurabilities,
  and  as illustrated in this paper, it is not necessary 
 to assume that the planets formed beyond an ice line and then underwent very large changes to their semi-major axes.

In this paper we study the evolution of   multi-planet  systems  with  masses 
that are small enough that the tidal interaction with the disk, with the exception of the estimation of coorbital torques
that can be important for migration and which are nearly always nonlinear (Paardekooper \& Papaloizou 2009),
is in the linear regime.

 We  develop a simplified  general analytic model describing the evolution of a  system of $N$ coplanar resonantly coupled 
 planets in near circular orbits with fixed orbital period ratios.  This enables rates of convergent migration  to be  estimated  given an orbital architecture,
 circularization times estimated from the theory of disk-planet tidal interaction, and the planet masses.
 This is applied to a system  with the current architecture of the Kepler 444 system.  As radii  are known but masses unknown for the planets in  that system, 
we work with a system which has  the same architecture but masses  determined such that estimated 
planetary  migration rates  are approximately  inversely proportional
 to their  masses,   as is  expected for disk-planet tidal interactions under  gravity.
  
We perform numerical simulations to study  the evolution of planetary systems  of the type described above,
as well as  systems with larger masses and systems  with consecutive pairs closer to resonance.
The analytic model is used as a guide and to provide input parameters.
The simulations  are carried out to investigate the maintenance and formation of comensurabilities
as the system migrates inwards significantly, changing its radial scale on a time scale
characteristic of the lifetime of current protoplanetary disks.
We find that significant convergent migration does not occur with migration rates estimated assuming a  system with a steady
architecture    corresponding to that of Kepler 444.  In order for this to occur,  consecutive pairs in  the system  need to have been 
closer to resonance during  an early phase when relative migration rates were faster. We speculate that
these rates could have slowed  as the system approached the inner edge of a truncated protoplanetary disk.  

The plan of the paper is as follows. In
Section \ref{sec3} we review relevant aspects of the theory of
orbital migration induced by the tidal interaction of protoplanets with protoplanetary disks, going on to give
estimates  of the orbital circularization times arising from both tidal interaction with the disk
and the central star  in  Sections \ref{circtime} and \ref{sec4}. In Section \ref{sec2} we set out the
basic equations  for a  model of a planetary system interacting with a protoplanetary disk
in which  it is  treated as $N$ interacting bodies. 

We then go on to formulate  the analytic model for a planetary system with consecutive pairs in resonance  
undergoing orbital migration and  circularization in Section \ref{sec5}.
In this formalism gravitational interaction only occurs between neighbouring planets
 through either one or two retained resonant angles.
 However, it is assumed that three body resonances of the Laplace type are absent.
The procedure may be applied to a system as a whole or to two or more uncoupled subsystems.
For a system or subsystem, with given masses,  in which  commensurabilities  are maintained, the model  leads to
relationships  between migration and circularization times that are  developed in  Section \ref{anmig}.
These enable  migration times for each planet  to be determined once planet masses, circularization times 
and the migration time for one of them is specified.
Migration times estimated  in  this way were used as input data for the detailed numerical simulations. 
 
Specification of the input  orbital parameters  and planet masses for the simulations is described in detail in Sections \ref{Simparam}-\ref{Planetmasses}.
How results may be scaled  so that the initial systems start at an expanded length scale and finish
in configuration resembling the initial one is indicated in 
Section \ref{Generalscaling}.  We go on to describe the results of the simulations in
Sections \ref {simres} - \ref{incloserr} and then summarize and discuss our results in Section \ref{discu}.

\section{Orbital migration and circularization due to tidal interaction with the protoplanetary disk}\label{sec3}
It is  well known that orbiting protoplanets  embedded in a protoplanetary disk
experience orbital migration and circularization as a result of tidal interaction
with it. (see eg. Ward 1997;  Papaloizou \& Terquem  2006; Correa-Otto et al. 2013,  Baruteau et al. 2014).
While the orbital circularization of low mass protoplanets may be robustly estimated
from a linear response calculation (see Section \ref{circtime} below) the situation regarding orbital
migration is less clear largely because of   significant  contributions  from 
nonlinear coorbital torques that depend on  conditions close to the planet 
as well as details of the protoplanetary disk model (e.g., Paardekooper \& Mellema 2006, Paardekooper \& Papaloizou 2009 ). It is well known that simple models of locally isothermal disks 
with surface density profiles that render coorbital torques ineffective produce inward migration times
for low mass protoplanets  that are too small by between one and two orders of magnitude (eg. Ida \& Lin 2008).

Various mechanisms for reducing or reversing  such migration have been proposed. 
These include the introduction of stochastic torques resulting from turbulence (Nelson \& Papaloizou 2004), invoking an eccentric disk (eg. Papaloizou 2002), 
the operation of coorbital torques induced by entropy gradients (eg. Paardekooper \& Mellema 2006).
the influence of outwardly  propagating density waves (eg. Podlewska-Gaca et al. 2012),
 the effects of a  magnetic field ( Terquem 2003, Guilet et al. 2013),   the influence of a turbulence driven  wind  on the disk  surface density profile  
 ( Suzuki et al. 2010, Ogihara et al. 2015 ),
 and more recently  the effects of heat radiation from an embedded protoplanet on an asymmetric disk mass distribution (Benitez-Llambay et al. 2015 ).
Some or all of these mechanisms may play a role in different regions of the protoplanetary disk
making the determination of migration rates problematic.

In the work presented here we  shall  instead consider the estimation of migration rates from
the  orbital configuration of a near resonant system together with
circularization times estimated from disk planet interaction, assuming that these can be related.
In general as was  found by Ida \& Lin (2008) for population synthesis modelling,
theoretical   migration rates determined for locally isothermal disk models  need to be significantly reduced to match these estimates. 
We suppose that one or more of the  mechanisms enumerated above act to provide the required reduction in the inner parts of the protoplanetary disk that we consider, 
However, for the low mass planets we consider,  we shall retain the expected approximate 
proportionality of the migration time to the reciprocal of the planet mass, 
as this is expected on very general grounds for purely gravitational  interaction
 ( eg. Papaloizou \& Terquem  2006;  Baruteau et al. 2014). But note that  this  could be departed from if effects due to heat radiation are  very significant as proposed by  Benitez-Llambay et al. (2015 ).
 Note that the circularization times discussed below are inversely proportional to the planet mass.

\subsection{Estimation  of  the orbital circularization time}\label{circtime}
The circularization time obtained from  a linear response calculation of the
disk-protoplanet interaction for a protoplanet of mass $m_j$  with small eccentricity  is given by (see Tanaka \& Ward 2004)
\begin{equation}
t_{e,j} =1.3\times10^3 \frac{M}{M_{\odot}}\frac{M}{m_{j}}\frac{\pi}{n_{j}}\left(\frac{H}{a_j}\right)^4\frac{M_J}{\pi \Sigma_p a_j^2}\label{t1circ}
\end{equation}
Here the semi-major axis of the planet is $a_j$ and its mean motion $n_j.$ 
The local disk semi-thickness is $H,$ the surface density at the position of the planet
is $\Sigma_p,$   the central mass is $M$ and $M_J$ is a Jovian mass. The disk is assumed to be locally isothermal.
In order to estimate $H$ we adopt the simple estimate given by
\begin{equation}
\frac{H}{a_j}= \sqrt{\frac{ {\cal R}T_j}{\mu n_j^2 a_j^2}}
\end{equation}
where the disk temperature at  the location of the planet
is $T_j= T_{eff}\sqrt{R_*/2a_j}.$ Here the mean molecular weight is
$\mu,$  $T_{eff}$ is the effective temperature of the central star, $R_*$ is its radius
and ${\cal R}$ is the gas constant.
As an ilustration we  adopt the stellar parameters to be those for Kepler 444 (Campante et al. 2015),
 and $\mu=2.$ 
Then we obtain
\begin{equation}
\frac{H}{a_j}=  0.02 \left(\frac{  a_j}{0.1AU}\right)^{1/4}.
\end{equation}
For aspect ratios of the above estimated magnitude  and the planet masses we consider, the disk planet interaction is expected to be in the linear
type I migration regime (eg. Ward 1997), justifying the use of equation  (\ref{t1circ}), 
from which we obtain
\begin{equation}
t_{e,j} =1.6\times10^{-4}\frac{M}{m_{j}}\frac{\pi}{n_{j}}\frac{a_j}{0.1AU}\frac{M_J}{\pi \Sigma_p a_j^2}.
\end{equation}
We note that for the special case with $\Sigma_p \propto a_j^{-1},$ $t_c \propto n_j^{-1} $ scales as the orbital period.
Then we have
\begin{equation}
t_{e,j} =3.2\times10^{-4}\frac{M}{m_{j}}\frac{\pi}{n_{j}}\left(\frac{a_j}{0.1AU}\frac{M_J}{M_D(a_j)}\right)\label{circmodele}.
\end{equation}
In the above expression $M_D(a_j)$ is the disk mass interior to $a_j$ and the quantity in brackets is constant.
Thus for $M_D = M_J $  interior to  $1 AU,$ 
we obtain $t_c \sim 360$ orbital periods for a one earth mass protoplanet and $M= 0.758M_{\oplus}.$ 
We note that if the above mass scaling applies out to $5AU,$ the interior disk mass would
correspond to  two and a half times  that of the minimum mass solar nebula.
However, we stress that we are only concerned with the dynamics interior to $1AU$
and this would be unaffected if the disk surface density was decreased at larger radii
so that such an extrapolation does not apply.

\subsection {Orbital circularization due to tides from the central star}\label{sec4}
The circularization timescale due to tidal interaction with the star
was obtained from   Goldreich~\& Soter (1966) in the form
\begin{equation}
t_{e,j}^s = \frac{0.69 \rho_{pj}^{5/3} a_{j}^{13/2} Q'}{G^{1/2}M^{3/2}m_{j}^{2/3}} 
\label{teccs}
\end{equation}
\noindent where $R_{pj}$ is the radius of planet $j$  and  $\rho_{pj}$ is its mean density. The quantity
$Q'= 3Q/(2k_2),$ where $Q$ is the tidal dissipation function and $k_2$
is the Love number.  
 The values of these  tidal parameters  applicable to  exoplanets  are very  uncertain. 
 However, for solar system planets in the terrestrial mass
range, Goldreich \& Soter (1966) give estimates for $Q$ in the range
10--500 and $k_2 \sim 0.3$, which correspond to $Q'$ in the range
50--2500.  We may also write (\ref{teccs}) in the form
\begin{equation}
t_{e,j}^s = \frac{9.6\times10^5 (\rho_{pj}/(1gmcm^{-3}))^{5/3}( a_{j}/0.04AU)^{13/2}( Q'/100) }{(m_{j}/M_{\oplus})^{2/3} (M/M_{\oplus})^{3/2}} y. 
\label{teccsn}
\end{equation}
From this we see that sub earth mass rocky planets orbiting further out than $0.04AU$
such as those in Kepler 444 are unlikely to  more than barely circularize  due to this mechanism within an expected formation time of $10^{6-7} y.$ 
It is  accordingly far less effective than interaction with the disk during this period
so we shall neglect it. However, it may result in a small  rate of separation of the system away from resonance
after disk dispersal (eg. Papaloizou 2011).
 We note that  convergent  disk  migration  of consecutive pairs of planets leads naturally to  multiple systems
in resonant chains of the type considered here (eg. Cresswell \& Nelson 2006;  Papaloizou \& Terquem 2010;  Baruteau et al. 2014).
Note that if they start out close to resonance,  these  configurations may be produced with  the system as a whole undergoing   little net radial migration.

\section{Basic equations for an N body model of a planetary system interacting with a disk }\label{sec2}
We begin by
considering  a system of $N$ planets and a central star interacting gravitationally.
The equations of motion are:
\begin{equation}
{d^2 {\bf r}_j\over dt^2} = -{GM{\bf r}_j\over |{\bf r}_j |^3}
-\sum_{k=1\ne j}^N {Gm_k  \left({\bf r}_j-{\bf r}_k \right) \over |{\bf
    r}_j-{\bf r}_k |^3} -{\bf \Gamma} -{\bf \Gamma}_{j} -{\bf B}_{j} \; ,
\label{emot}
\end{equation}

\noindent  where $M$,  $m_j,$  and ${\bf r}_j$ denote the mass of
the central star,  the mass of planet~$j,$   and the position vector of planet
$j,$ respectively.  The acceleration of the coordinate system based on
the central star (indirect term) is:
\begin{equation}
{\bf \Gamma}= \sum_{j=1}^N {Gm_j{\bf r}_{j} \over |{\bf r}_{j}|^3},
\label{indt}
\end{equation}

\noindent Orbital circularization due to  tidal interaction with  the disk
 is dealt with through the addition of a  frictional  damping force taking the form (see eg.
Papaloizou~\& Terquem~2010)
\begin{equation}
{\bf \Gamma}_{j} =
 - \frac{2}{|{\bf r}_j|^2 t_{e,j}} \left( \frac{d {\bf r}_j}{dt} \cdot
{\bf r}_j \right) {\bf r}_j 
\label{Gammai}
\end{equation}
\noindent  Migration torques
 are included through ${\bf B}_{j}$ which takes the form ( see eg. Terquem~\&
Papaloizou~2007)
\begin{equation}
{\bf B}_{j} =
 - \frac{{\bf r}_j\times\left( {\bf r}_j\times (d {\bf r}_j/dt) \right) }{3 t_{mig,j}|{\bf r}_j|^2},  
\label{Bi}
\end{equation}
where $t_{mig,j}$ is  defined to be the migration time for planet $j,$
being the characteristic time on which the mean motion increases. 
 Note that the characteristic time on which the specific angular momentum
decreases is $3t_{mig,j}.$ Hence the factor of 3 in the denominator of 
equation (\ref{Bi}).

\section{Analytic model for a planetary system with consecutive pairs in resonance  
undergoing orbital migration and  circularization}\label{sec5}

We  develop an analytic model that shows how a system of $N$
planets   undergoes orbital evolution driven by orbital migration and  circularization
driven by tidal interaction with a protoplanetary disk.
The coupling between the planets occurs through the behaviour of resonant angles, which may librate,
so  producing  evolution of orbital elements after time averaging.
  However, significant
deviations from exact commensurability may occur.  
We begin by formulating equations governing the system
without forces associated with disk planet interaction. 
In that case a Hamiltonian formalism can be used. 
We then go on to add effects arising from interaction with the disk.
The starting point is the same as in Papaloizou (2015).
However, in contrast to the work there, the discussion here includes the effects of orbital migration
and considers a system with an arbitrary number of planets.

\subsection{Coordinate system}
We adopt Jacobi coordinates (eg. Sinclair~1975 ) for which the radius vector
 of  planet $j,$  ${\bf r }_j,$ is measured relative
to the  centre of mass of  the system comprised of $M$ and  all other  planets
interior to  $j,$ for  $j=1,2, 3, ..., N.$ Here  $j=1$ corresponds to the
innermost planet and $j=N$ to the outermost planet.
 
 \subsection {Hamiltonian for the system without disk interaction}\label{Hamilsec}
The  Hamiltonian for the system governed by (\ref{emot}) with orbital migration and circularization absent can be written,  correct to second order
in the planetary masses,  in the form:
\begin{eqnarray} H & = &  \sum_{j=1}^N \left({1\over 2}  m_j | \dot {\bf r}_j|^2
- {GM_{j}m_j\over  | {\bf r}_j|} \right)   \nonumber \\
& - &\sum_{j=1}^N\sum_{k=j+1}^NGm_{j}m_k
\left({1 \over  | {\bf r}_{jk}|}  -  { {\bf r}_j\cdot {\bf r}_k
\over  | {\bf r}_{k}|^3}\right).
\end{eqnarray}
Here $M_{j}=M+m_j $ and
$ {\bf r}_{jk}= {\bf r}_{j}- {\bf r}_{k}.$

Assuming,  the planetary system is strictly coplanar, the equations governing the  motion  
 about a dominant central mass,
 may be written in the form
(see, e.g., Papaloizou~2003; Papaloizou \& Szuszkiewicz~2005, Papaloizou~2015):

\begin{eqnarray}
\dot E_j &=& -n_j\frac{\partial H}{\partial \lambda_j}\label{eqnmo1}\\
\dot L_j &=& -\left(\frac{\partial H}{\partial \lambda_j}+\frac{\partial H}{\partial \varpi_j}\right)\\
\dot \lambda_j &=& \frac{\partial H}{\partial L_j} + n_j \frac{\partial H}{\partial E_j}\\
\dot \varpi_j &=& \frac{\partial H}{\partial L_j}.\label{eqnmo4}
\end{eqnarray}

Here  and in what  follows unless stated otherwise, 
 $m_j$ is replaced by he reduced mass  so that $m_j  \rightarrow m_{j}M/(M+m_{j}).$ 
The orbital  angular momentum of  planet  $j$ 
is $L_j$ and the
orbital  energy is $E_j.$
 The  mean longitude of planet $j$ is $\lambda_j = n_j (t-t_{0j}) +\varpi_j ,$
 with  $n_j  = \sqrt{GM_{j}/a_j^3}= 2\pi/P_j $ being  the  mean motion,  and
$t_{0j}$ denoting the time of periastron passage.  The semi-major axis and orbital period   
 of planet  $j$ are  $a_j$  and $P_j.$   
 The longitude of periastron is $\varpi_j.$
 The quantities $\lambda_j,$ $\varpi_j,$ $L_j$ and $E_j$ can be used to describe the dynamical
 system described above.

\noindent However,  we note that for motion around a central point  mass $M$ we have:
\begin{eqnarray}
     L_j &=&  m_{j}\sqrt{GM_{j}a_i(1-e_j^2)}, \\
     E_j &=& -{{GM_{j}m_{j}}\over{2a_j}},
\end{eqnarray}
where $e_j$  the eccentricity of planet $j.$ 
By making use of these relations  we  may adopt 
$\lambda_j,$ $\varpi_j,$ $a_j$ or equivalently $n_j,$  and $e_j$ as dynamical variables. 
 We comment that the difference between  taking $m_j$
to be the reduced mass rather than  the actual mass  of planet $j$ when evaluating $M_j$  in the  expressions for
 $L_j$ and $E_j$
is third order in the typical planet to star mass ratio and thus it may be neglected.
The equations we  ultimately  use  turn out to be the same as those obtained assuming the central mass is fixed. 
 The Hamiltonian may quite generally
 be expanded in a Fourier series
involving linear combinations of the $2N-1$   angular differences
$\varpi_i -\varpi_1,$ $i=2, 3, ...,  N$ and
$\lambda_i - \varpi_i,  i=1,2,3, ..., N. $ 

 Here we are interested
in the effects of the $N-1$  first order $p_i+1: p_i, $  $i= 1,2,3, ... , N-1,$    commensurabilities
associated with the planets  with masses  $m_i$ and $m_{i+1}$ respectively.
In this situation,  we expect that any of the $2(N-1)$  angles
$\Phi_{i+1,i,1} = (p_i+1)\lambda_{i+1}-p_i\lambda_i-\varpi_i, $ 
$\Phi_{i+1,i,2} = (p_{i}+1)\lambda_{i+1}-p_i\lambda_{i}-\varpi_{i+1},$ $i=1,2,3, ...,  N-1,$
may  be slowly varying.
 Following  standard practice
 (see, e.g.,  Papaloizou \& Szuszkiewicz~2005; Papaloizou \& Terquem~2010),
high frequency terms in the Hamiltonian are averaged out.
In this way,  only terms in the Fourier expansion involving  linear
combinations of $\Phi_{i+1,i,1},$  and $\Phi_{i+1,i,2},$ $i=1,2,3, ..., N-1,$
as argument are  retained. 

Working in the limit of small eccentricities that is applicable here,
 terms that are higher order than first  in the eccentricities can also be  discarded.
The  Hamiltonian   may  then be written
in the form:
\begin{equation} H=\sum_{k=1}^{k=N}E_k +  \sum_{k=1}^{N-1}  H _{k,k+1} ,\label{Hamil0} \end{equation}
 where:
\begin{equation} \hspace{-1mm} H _{k, k+1}= -\frac{Gm_km_{k+1}}{a_{k+1}}\left[  e_{k+1}C_{k,k+1}\cos \Phi_{k+1,k,2}
- e_kD_{k,k+1}\cos\Phi_{k+1, k, 1} \right] \label{Hamil} \end{equation}
with:
\begin{eqnarray} \hspace{-6mm}C _{i,j} & = & {1 \over 2}\left(   x_{i,j}{db^{m}_{1/2}(x)\over dx} \Biggl|_{x=x_{i,j}}\Biggr. +(2m+1)b^{m}_{1/2}(x_{i,j})
-(2m+2)x_{i,j}\delta_{m,1} \right),   \label{Hamil1} \\
\hspace{-6mm}D _{i,j} & =& {1 \over 2}\left(   x_{i,j}{db^{m+1}_{1/2}(x)\over dx}\Biggl|_{x=x_{i,j}}\Biggr. +2(m+1)b^{m+1}_{1/2}(x_{i,j})  \right) . \label{Hamil2} 
\end{eqnarray}
Here the integer $m=p_i$ 
 and  $b^{m}_{1/2}(x)$ denotes  the usual Laplace coefficient
(e.g., Brouwer \& Clemence 1961;  Murray and Dermott 1999)
with the argument $x_{i,j} = a_i/a_j.$

\subsection{Incorporation of orbital  migration and circularization due to interaction with the protoplanetary disk}\label{sec4.3}
 
 Using equations~(\ref{eqnmo1})--(\ref{eqnmo4})
together with equation~(\ref{Hamil0})  we may first  obtain the equations of motion
without the effect of  migration and circularization  due to   interaction with the protoplanetary disk. 
Having obtained  the former, the effect of the latter  may be added in (see eg. Papaloizou~2003).
 Following this procedure,  we obtain:
\begin{align}
&\hspace{-.3cm}\frac{d e_j}{dt}=\frac{n_j}{M_j}\left[ m_{j+1}  \frac{a_j}{a_{j+1}}
D_{j,j+1}\sin\Phi_{j+1, j, 1}                                
- m_{j-1} C_{j-1,j}\sin \Phi_{j,j-1,2}\right]  - \frac{e_j}{t_{e,j}}, \label{eqntid1}\\
\!\!\!\!\!\!\!\!\!\!\!
&\hspace{-.3cm} \frac{d n_j}{dt} =
\frac{3(p_{j-1}+1)n_j^2m_{j-1}}{M_j}\left(C_{j-1,j}e_j\sin \Phi_{j,j-1,2}-D_{j-1,j }e_{j-1}\sin \Phi_{j,j-1,1} \right)\nonumber\\
   &\hspace{-.4cm}-
\frac{3p_j n_j^2m_{j+1}a_j}{M_ja_{j+1}}\left(C_{j,j+1}e_{j+1}\sin\Phi_{j+1,j,2}-D_{j,j+1}e_j\sin\Phi_{j+1,j,1} \right)\nonumber\\
&\hspace{-.3cm}+\frac{3n_je_j^2}{t_{e,j}}   +\frac{n_j}{t_{mig,j}} , \label{eqntid2} \\
\!\!\!\!\!\!\!\!\!\!
&\vspace{0.4cm}\nonumber\\
\!\!\!\!\!\!\!\!\!\!
&\hspace{-.2cm}
{\rm for {\hspace{2mm} }}j = 1,2,3,4, ..., N\hspace{6mm} {\rm and}\nonumber \\
\!\!\!\!\!\!\!\!\!\!
&\vspace{0.5cm}\nonumber\\
&\hspace{-.3cm}\frac{d  \Phi_{j+1,j,1}}{dt} = (p_j+1)n_{j+1}-p_jn_j\nonumber\\
&\hspace{-.3cm}-\frac{n_j}{e_j}\left[ \frac{m_{j-1}}{M_j} C_{j-1,j}\cos \Phi_{j,j-1,2}-\frac{m_{j+1}a_{j}}{M_{j} a_{j+1}}D_{j,j+1}\cos \Phi_{{j+1,j,1}}\right]   ,
\hspace{2mm}{\rm with}  \label{eqntid3}\\
&\hspace{-.3cm}\frac{d  \Phi_{j+1,j,2}}{dt } = (p_j+1)n_{j+1}-p_jn_j\nonumber\\
&\hspace{-.3cm}-\frac{n_{j+1}}{e_{j+1}}\left[ \frac{m_{j}}{M_{j+1}} C_{j,j+1}\cos \Phi_{j+1,j,2}-\frac{m_{j+2}a_{j+1}}{M_{j+1} a_{j+2}}D_{j+1,j+2}\cos \Phi_{{j+2,j+1,1}}\right] .
 \label{eqntid4}\\
 \!\!\!\!\!\!\!\!\!\!
 \!\!\!\!\!\!\!\!\!\!
& \vspace{4mm}\nonumber\\
&\hspace{-.2cm}
{\rm for {\hspace{2mm} }}j = 1,2,3, ... ,N-1 \hspace{6mm} \nonumber \\
\nonumber
\end{align}






\noindent We remark that terms on the right hand sides of the above equations for  which $j$  takes on a value such that  a factor  $m_0$ or $m_{N+1}$ is implied 
are to be omitted ( or one may set $m_0 = m_{N+1} =0$).

\noindent   At this point we note from   
 equations    (\ref{eqntid3}) and (\ref {eqntid4}) that 
 \begin{align}
 &\hspace{-1.3cm} \frac{d  \Phi_{j+1,j,1}}{dt} -\frac{d  \Phi_{j,j-1,2}}{dt}= (p_j+1)n_{j+1}- (p_{j}+p_{j-1}+1)n_{j}+p_{j-1}n_{j-1}\label{sec4.3eq}\\
 &\hspace{-1.3cm} {\rm for} \hspace{2mm} j=2,3, ..., N-1.\nonumber
 \end{align}
 Thus if both angles $\Phi_{k+1,k,1}$  and  $\Phi_{k,k-1,2}$ are such that the time average of their derivatives is zero, then
the  the time averages of the Laplace resonance~relations~\begin{align}
& \hspace{-4cm}(p_k+1)n_{k+1}- (p_{k}+p_{k-1}+1)n_{k}+p_{k- 1}n_{k-1} =  0\label{LaLa}\\
&\hspace{-4cm}{\rm for}  {\hspace{2mm} } k  = 2,3, ..., N-1 \nonumber
\end{align}~are satisfied. A special case is when the angles are constant in which case the Laplace resonance relations are satisfied
without the need of a time average.\\
 For a system of $N$ planets there could be up to $N-2$ Laplace resonance relations.  If  none of these exist after time averaging
then a maximum of $2N -2 -(N-2)=N$ angles may be librating. We focus on the case when this is the situation.
One can see that in order for the planets to be interacting with non zero eccentricities, at least one resonant angle associated with each consecutive
pair must contribute and then both angles can librate for only one consecutive pair.
We suppose that this pair corresponds to $k = N - J$ and $k = N - J +1$ respectively for some integer $J.$
 Then in order that  all of the planets are involved in the  interacting system, the angles that may be  librating,
or more generally contributing to long term time averages have to be
$\Phi_{k+1,k,1} ,$   for $k =1,2, ..., N-J$ and $\Phi_{N-k+1,N-k,2},$  for  $k =1, 2, ..., J.$
Retaining only  these angles from now on,  equations (\ref{eqntid1}) - (\ref{eqntid4}) then provide $3N$ equations  for the $3N$ quantities
comprising these angles and  the quantities $n_k, e_k, k=1,2,..N.$  These take the form

\begin{align}
&\hspace{-0.5cm}\frac{d e_j}{dt}=\frac{n_j}{M_j} m_{j+1}  \frac{a_j}{a_{j+1}}
D_{j,j+1}\sin(\Phi_{j+1, j, 1})                                
 - \frac{e_j}{t_{e,j}} \label{eqntid5}\\
&\hspace{-0.5cm}
{\rm for {\hspace{2mm} }} j = 1,2,3, ..., N-J, \hspace{6mm} \nonumber \\
&\vspace{0.1mm}\nonumber\\
&\hspace{-0.5cm}\frac{d e_{N-j+1}}{dt}=-\frac{n_{N-j+1}}{M_{N-j+1}}
m_{N-j} C_{N-j,N-j+1}\sin (\Phi_{N-j+1,N-j,2})  - \frac{e_{N-j+1}}{t_{e,N-j+1}}  \label{eqntid5N}\\
&\hspace{-0.4cm}{\rm for {\hspace{2mm} }} j = 1,2,3, ..., J,\nonumber\\ 
\!\!\!\!\!\!\!\!\!\!\!\
&\vspace{0.1mm}\nonumber\\
&\hspace{-0.5cm} \frac{d n_j}{dt} = \frac{3n_je_j^2}{t_{e,j}} +\frac{n_j}{t_{mig,j }}
-\frac{3(p_{j-1}+1)n_j^2m_{j-1}}{M_j}D_{j-1,j }e_{j-1}\sin \Phi_{j,j-1,1}\nonumber\\ 
 & +
\frac{3p_j n_j^2m_{j+1}a_j}{M_ja_{j+1}}\left( D_{j,j+1}e_j\sin\Phi_{j+1,j,1}- \delta_{j,N-J}C_{j,j+1}e_{j+1}\sin\Phi_{j+1,j,2}\right) 
\label{eqntid6} \\
&\hspace{-0.4cm}{\rm for {\hspace{2mm} }}j = 1,2,3, ..., N-J, \hspace{6mm} \nonumber \\
&\vspace{0.1mm}\nonumber\\
&\hspace{-0.5cm} \frac{d n_{N-j+1}}{dt} =
\frac{3(p_{N-j}+1)n_{N-j+1}^2m_{N-j}}{M_{N-j+1}}\left(C_{N-j,N-j+1}e_{N-j+1}\sin \Phi_{N-j+1,N-j,2}\right.\nonumber\\ 
&\hspace{-0.5cm}\left. -\delta_{j,J}D_{N-j,N-j+1 }e_{N-j}\sin \Phi_{N-j+1,N-j,1}\right)\nonumber\\
& \hspace{-0.5cm}-\frac{3p_{N-j+1}n_{N-j+1}^2m_{N-j+2}a_{N-j+1}}{M_{N-j+1}a_{N-j+2}} \left. C_{N-j+1,N-j+2}e_{N-j+2}\sin \Phi_{N-j+2,N-j+1,2}\right. \nonumber\\ 
&\hspace{-0.5cm}+\frac{3n_{N-j+1}e_{N-j+1}^2}{t_{e,N-j+1}}+\frac{n_{N-j+1}}{t_{mig,N-j+1}}  \label{eqntid6N} \\
 &\hspace{-0.5cm}{\rm for {\hspace{2mm} }} j = 1,2,3, ..., J, \hspace{2mm} {\rm and} \nonumber\\
&\vspace{0.1mm}\nonumber\\
&\hspace{-0.5cm} \frac{d  \Phi_{j+1,j,1}}{dt}= (p_j+1)n_{j+1}-p_jn_j
+\frac{n_j}{e_j}\frac{m_{j+1}a_{j}}{M_{j} a_{j+1}}D_{j,j+1}\cos \Phi_{{j+1,j,1}}   
 \label{eqntid7}\\
 &\hspace{-0.5cm}
{\rm for {\hspace{2mm} }}j = 1,2,3, ..., N-J, \hspace{2mm} {\rm with} \hspace{6mm} \nonumber \\
&\vspace{0.1mm}\nonumber\\
&\hspace{-0.5cm}\frac{d  \Phi_{N-j+1,N-j,2}}{dt } = (p_{N-j}+1)n_{N-j+1}-p_{N-j}n_{N-j} \hspace{2mm} \nonumber\\
&\hspace{-0.5cm}-\frac{n_{N-j+1} m_{N-j}  C_{N-j,N-j+1}}{e_{N-j+1} M_{N-j+1} } \cos \Phi_{N-j+1,N-j,2}\nonumber \\ 
 &\hspace{-0.5cm}{\rm for {\hspace{2mm} }} j = 1,2,3, ..., J. \nonumber
 \label{eqntid7N}\\
\end{align}
Here, as above, as terms involving an implied $m_0$ or $m_{N+1}$ should be absent. Thus they should be set to  zero.
\subsubsection{Division into subsystems}\label{subsystems}
We further remark that the above discussion applies to an entire system of $N$ planets.
Instead of this it is possible to split the system into two or more non interacting subsystems
to each of which the above discussion applies. For the example of a five planet system that will be considered below,
we could apply the analysis to the whole system with for example $J=2.$ Then all consecutive pairs are coupled by single first order resonances 
 except planets $3$ and $4$ which are coupled by a pair of first order resonances.
 
  Alternatively we could consider
the innermost pair as a separate system to that of the outermost three.  Then both the innermost pair and planets $3$ and $4$
could be coupled by  pairs of first order resonances with the outermost pair coupled by a first order resonance. This case is dealt with using the same analysis
but applied to two sub systems the first with  $N=2,$ and $J=1$ and the second with $N=3$ and $J=2.$
When this is done one migration time needs to be specified for each subsystem. We took these  to be for  the innermost planets.

\subsubsection{Auxiliary quantities}
In what follows below we  shall find it  useful to define the auxiliary quantities $x_j = e_j \cos \Phi_{j+1,j,1} $ and $y_j = e_j \sin \Phi_{j+1,j,1}. $
From equations (\ref{eqntid5}) and (\ref{eqntid7}) we find a  pair of equations in terms of these quantities
in the form

\begin{align}
&\hspace{-0.5cm} \frac{d  x_{j}}{dt}= -\left( (p_j+1)n_{j+1}-p_jn_j\right)y_j  -\frac{x_j}{t_{e,j}} \hspace{6mm}{\rm and}\label{eqntid10}\\
&\hspace{-0.5cm} \frac{d  y_{j}}{dt}= \left( (p_j+1)n_{j+1}-p_jn_j\right)x_j 
+n_j\frac{m_{j+1}a_{j}}{M_{j} a_{j+1}}D_{j,j+1}  -\frac{y_j}{t_{e,j}}, \label{eqntid11}\\
 &\hspace{-0.5cm}{\rm for {\hspace{2mm} }}j = 1,2,3, ..., N-J.\hspace{6mm} \nonumber \
\end{align}
\noindent Similarly for  $j = 1,2...J, $ we define the auxiliary quantities

\noindent  $x_{N-j+1} = e_{N-j+1} \cos \Phi_{N-j+1,N-j,2} $ and $y_{N-j+1} = e_{N-j+1} \sin \Phi_{N-j+1,N-j,2}. $
From equations (\ref{eqntid5N}) and (\ref{eqntid7N}) we find a  pair of equations in terms of these quantities
in the form

\begin{align}
&\hspace{-0.5cm} \frac{d  x_{N-j+1}}{dt}= -\left( (p_{N-j}+1)n_{N-j+1}-p_{N-j}n_{N-j}\right)y_{N-j+1}  -\frac{x_{N-j+1}}{t_{e,N-j+1}} \label{eqntid12}\\
&{\rm and}\nonumber\\
&\hspace{-0.5cm} \frac{d  y_{N-j+1}}{dt}= \left( (p_{N-j}+1)n_{N-j+1}-p_{N-j} n_{N-j}\right)x_{N-j+1}\nonumber\\
&-n_{N-j+1}\frac{m_{N-j}}{M_{N-j+1} }C_{N-j,N-j+1}    -\frac{y_{N-j+1}}{t_{e,N-j+1}}.
 \label{eqntid13}
\end{align}

\noindent  At this point we note that in the analysis below we 
use equations    (\ref{eqntid1}) - (\ref {eqntid4}) to
calculate  perturbations to the  orbital elements and resonant angles
correct to first order in the typical planet to central star mass ratio.
For this purpose from now on  we  adopt the actual mass  of planet $i$ for $m_{i}, $ rather than the reduced mass and replace $M_i$ by $M,$
as any consequent   corrections  will be second order.  Similarly the difference between using 
 actual or reduced masses when evaluating the circularization  times may be neglected as any corrections to those will also be correspondingly small.
Accordingly $m_i$ will be identified as being the actual mass of planet $i$ everywhere  from now on. 

\noindent The terms involving  the circularization times
$t_{e,i}$  are associated with effects  arising from  
 interaction of  planet $i$ with the protoplanetary disk.
We shall assume throughout that such terms, though being retained, can be  of lower order than those proportional to the disturbing masses, $m_i.$
However, we shall assume that expressions that are of second or higher order in   these quantities  may be neglected.

We consider  solutions of (\ref{eqntid5}) - (\ref{eqntid7N}) for which the time average of the rates of change of $e_j,$ $x_j,$ $y_j,$ $j= 1,2, ..., N$ 
and the retained resonant angles  is zero. Performing a time average of 
equations (\ref{eqntid5}) and (\ref{eqntid5N}) then 
implies  that 
\begin{align}
&\hspace{-0.5cm}  \frac{n_j}{M} m_{j+1}  \frac{a_j}{a_{j+1}}
D_{j,j+1}\langle \sin(\Phi_{j+1, j, 1})\rangle =       
  \frac{\langle e_j\rangle }{t_{e,j}} ,\label{eqntid50}\\
&\hspace{-0.5cm}
{\rm for {\hspace{2mm} }} j = 1,2,3, ..., N-J \hspace{6mm}{\rm and}  \nonumber \\
&\hspace{-0.5cm}\frac{n_{N-j+1}}{M}
m_{N-j} C_{N-j,N-j+1}\langle \sin (\Phi_{N-j+1,N-j,2}) \rangle  =- \frac{\langle e_{N-j+1}\rangle }{t_{e,N-j+1}}, \label{eqntid50N}\\
&\hspace{-0.5cm}
{\rm for {\hspace{2mm} }} j = 1,2,3, ..., J.\nonumber
\end{align}
where the angled brackets denote a time average.
We also assume that the semi-major axes or equivalently the mean motions,  as well as their time derivatives,     
vary on a time scale that is long compared to
the  characteristic time over which averages are taken. Thus they are treated as being constant during the averaging 
procedure.


\noindent Multiplying  equation (\ref{eqntid5}) by $e_j$  and (\ref{eqntid5N}) by $e_N$ 
and performing a time average under the same assumptions yields in addition  
\begin{align}
&\hspace{-0.5cm}  \frac{n_j}{M} m_{j+1}  \frac{a_j}{a_{j+1}}
D_{j,j+1}\langle y_j\rangle =       
  \frac{\langle e_j^2\rangle }{t_{e,j}} ,\label{eqntid501}\\
&\hspace{-0.5cm}
{\rm for {\hspace{2mm} }} j = 1,2,3, ..., N-J \hspace{6mm}{\rm and}  \nonumber \\
&\hspace{-0.5cm}\frac{n_{N-j+1}}{M}
m_{N-j} C_{N-j,N-j+1}\langle y_{N-j+1} \rangle  = - \frac{\langle e_{N-j+1}^2 \rangle }{t_{e,N-j+1}}, \label{eqntid501N}\\
&\hspace{-0.5cm}
{\rm for {\hspace{2mm} }} j = 1,2,3, ..., J.\nonumber
\end{align}

The above expressions can be substituted into 
the time averaged form of equations (\ref{eqntid6}) and (\ref{eqntid6N}) for the rate of change on the mean motions.These are then found to be determined by the time averaged squares of the eccentricities,  
the circularization rates  and the migration rates according  to 

\begin{align}
&\hspace{-0.5cm} \frac{d n_j}{dt} =
{3(p_j+1) n_j} \frac{\langle e_j^2\rangle }{t_{e,j}} 
 +\frac{n_j}{t_{mig,j }}
-\frac{3(p_{j-1}+1)n_j^2}{n_{j-1}}\frac{m_{j-1}a_{j}}{m_{j}a_{j-1}}\frac{\langle e_{j-1}^2\rangle }{t_{e,j-1}}\nonumber\\
 &\hspace{-0.5cm} +\delta_{j,N-J}\frac{3p_{N-J}n_{N-J}^2}{n_{N-J+1}}\frac{m_{N-J+1}a_{N-J}}{m_{N-J}a_{N-J+1}}\frac{\langle e_{N-J+1}^2\rangle }{t_{e,N-J+1}}
  \label{eqntid66} \\
&\hspace{-0.5cm}
{\rm for {\hspace{2mm} }}j = 1,2,3, ..., N-J \hspace{6mm} \nonumber \\
&\vspace{2mm}\hspace{-5mm} {\rm and} \nonumber\\
&\hspace{-0.5cm} \frac{d n_{N-j+1}}{dt} =
-3p_{N-j}n_{N-j+1}\frac{\langle e_{N-j+1}\rangle^2}{t_{e,N-j+1}} \nonumber\\ 
&\hspace{-0.5cm}+\frac{3p_{N-j+1}n_{N-j+1}^2}{n_{N-j+2}}\frac{m_{N-j+2}a_{N-j+1}}{m_{N-j+1}a_{N-j+2}}\frac{\langle e_{N-j+2}^2\rangle }{t_{e,N-j+2}}\nonumber\\ 
&\hspace{-0.5cm}-\delta_{j,J}\frac{3(p_{N-j}+1)n_{N-j+1}^2}{n_{N-j}}\frac{m_{N-j}a_{N-j+1}}{m_{N-j+1}a_{N-j}}\frac{\langle e_{N-j}^2\rangle }{t_{e,N-j}}
+\frac{n_{N-j+1}}{t_{mig,N-j+1}} \label{eqntid66N}\\ 
&\hspace{-0.5cm}{\rm for {\hspace{2mm} }} j = 1,2,3, ..., J.\nonumber
\end{align}
or equivalently

\begin{align}
&\hspace{-0.5cm} \frac{d n_{j}}{dt} =
-3p_{j-1}n_{j}\frac{\langle e_{j}\rangle^2}{t_{e,j}} + \frac{3p_{j}n_{j}^2}{n_{j+1}}\frac{m_{j+1}a_{j}}{m_{j}a_{j+1}}\frac{\langle e_{j+1}^2\rangle }{t_{e,j+1}}\nonumber\\ 
-&\delta_{N-J+1,j}\frac{3(p_{j-1}+1)n_{j}^2}{n_{j-1}}\frac{m_{j-1}a_{j}}{m_{j}a_{j-1}}\frac{\langle e_{j-1}^2\rangle }{t_{e,j-1}}
+\frac{n_{j}}{t_{mig,j}}, \label{eqntid66N1}\\ 
&\hspace{-0.5cm}{\rm for {\hspace{2mm} }}  j  =  N-J+1, N-J+2,  N-J+3, ..., N.\nonumber    
\end{align}

\subsection{Relation between migration and circularization times for a system maintaining commensurabilities}\label{anmig}

\noindent If the system evolves maintaining commensurabilites,  the mean square eccentricities can be  related to the migration and circularization rates from 
$N-1$ equations found by setting $dn_{j+1}/d n_j = n_{j+1}/ n_j \sim  p_j/(p_j+1),$ for $j= 1,2,3,  ..., N-1$
These may be written
\begin{align}
&\hspace{-0.5cm} \frac{1}{t_{mig,j+1 }}  -\frac{1}{t_{mig,j }} = F_j -F_{j+1}\label{migpres}\\ 
\nonumber
\end{align}
where
\begin{align}
&\hspace{-0.5cm} F_j = -\frac{3(p_{j-1}+1)n_j}{n_{j-1}}\frac{m_{j-1}a_{j}}{m_{j}a_{j-1}}\frac{\langle e_{j-1}^2\rangle }{t_{e,j-1}}\nonumber\\
 &\hspace{-0.5cm} +\delta_{j,N-J}\frac{3p_{j}n_{j}}{n_{j+1}}\frac{m_{j+1}a_{j}}{m_{j}a_{j+1}}\frac{\langle e_{j+1}^2\rangle }{t_{e,j+1}}
  + {3(p_j+1) } \frac{\langle e_j^2\rangle }{t_{e,j}} \label{migpres1}\\
  &\hspace{-0.5cm}
{\rm for {\hspace{2mm} }}j = 1,2,3, ..., N-J, \hspace{6mm} \nonumber \\ \nonumber
\end{align}
and
\begin{align}
&\hspace{-0.5cm} F_j = -3p_{j-1}\frac{\langle e_{j}\rangle^2}{t_{e,j}} + \frac{3p_{j}n_{j}}{n_{j+1}}\frac{m_{j+1}a_{j}}{m_{j}a_{j+1}}\frac{\langle e_{j+1}^2\rangle }{t_{e,j+1}}\nonumber\\ 
&\hspace{-0.5cm}-\delta_{N-J+1,j}\frac{3(p_{j-1}+1)n_{j}}{n_{j-1}}\frac{m_{j-1}a_{j}}{m_{j}a_{j-1}}\frac{\langle e_{j-1}^2\rangle }{t_{e,j-1}}\label{migpres2}\\
  &\hspace{-0.5cm}
{\rm for {\hspace{2mm} }}j = N-J+1, N-J+2,  N-J+3, ..., N.   \hspace{6mm} \nonumber \\ \nonumber
\end{align}

\noindent  We also recall that we have adopted the convention that   terms  with implied $m_0$ or $m_{N+1}$ 
for $j=0$  and $j=N$ are  absent.

\noindent We remark that when the planets interact as considered here, it is not necessary that the rate of convergence
of every consecutive  pair in the absence of interaction be positive.   This is because the possible divergent  migration of either  component may
be blocked by interaction with  other neighbouring   components.
However, we can show that the rate of convergence of the innermost and outermost pair in the absence of interaction
must be positive as equations (\ref{migpres}) implies that

\begin{align}
&\hspace{-0.5cm} \frac{1}{t_{mig,N }}  -\frac{1}{t_{mig,1 }} = 
 \delta_{1,N-J}\frac{3p_{1}n_{1}}{n_{2}}\frac{m_{2}a_{1}}{m_{1}a_{2}}\frac{\langle e_{2}^2\rangle }{t_{e,2}}
  + {3(p_1+1) } \frac{\langle e_1^2\rangle }{t_{e,1}}
 \nonumber\\
&\hspace{-0.5cm}+3p_{N-1}\frac{\langle e_{N}\rangle^2}{t_{e,N}} +\delta_{N-J+1,N}\frac{3(p_{N-1}+1)n_{N}}{n_{N-1}}\frac{m_{N-1}a_{N}}{m_{N}a_{N-1}}
\frac{\langle e_{N-1}^2\rangle }{t_{e,N-1}}
 \label{eqntid67} 
\end{align}
  
\begin{table}
\begin{center}
     \begin{tabular}{| c |c | c | c | c | }
    \hline
    Run & Masses & Initial semi-major & Circularization & Migration \\
           &               &    axes  & times &   times \\ \hline
    $sacd $& $s $& $a$&$c$  & $d$\\ \hline
    $sacd_1$ &$ s$ & $a$&$c$ & $d_1$\\ \hline
   $ s_1ac_1f $& $s_1$ & $a$ & $c_1$& $f$ \\ \hline
    $s_1ac_1f_1$ & $s_1$ & $a$& $c_1$& $f_1$\\ \hline
      $sach$ &$ s$ &$ a$& $c$ &$ h$ \\ \hline
     $sacm $&$ s$& $a$&$c$ &   $m $\\ \hline
    $sbcm$ & $s$ &$ b$ &$ c$ & $m  $\\ \hline
        $sb_1cd_2$ &$ s$ & $b_1$& $c $ & $d_2$     \\ 
    \hline
  \end{tabular}
   {\caption{{\bf Table 1}  Parameters of the simulations are indicated
   The first column indicates the label defining the run.  The second  column indicates the planet masses adopted,  with $s$ denoting standard masses
   and $s_1$ denoting standard masses increased by a factor of ten. The third column  indicates the 
 initial semi-major axes (see  Section \ref{Migrationtimes}  for more details).
   The fourth  column indicates the circularization times used with $c$ denoting that the expression given by equation (\ref{circ}) was used and $c_1$ denotes values
   obtained from this expression increased by a factor $100.$  The final column indicates the migration times used (see Table 2 and Section \ref{Migrationtimes} for more details). }}
\end{center}
\end{table}


\begin{table}
\begin{center}
     \begin{tabular}{| c |c | c | c |c|  }
    \hline
    Migration time $\times 3n_j$ for planet $j$ & $d$ & $f$ & $h$ & $m$ \\ \hline
   $3n_1 t_{mig,1}/10^9$  & 3.247998 & 0.3247998&1.623999 & 0.9279994 \\ \hline
   $3 n_2t_{mig,2}/10^9$    & 2.569313  & 0.2569313  & 1.229178  & 0.6952312 \\ \hline
     $ 3n_3t_{mig,3}/10^9$&  3.214347  & 0.3214347  & 1.265322 &0.6938699   \\ \hline
     $3 n_4t_{mig,4}/10^9$& 1.313746  &0.1313746  &0.9268539  &0.5634547 \\ \hline
     $ 3n_5t_{mig,5}/10^9$ & 1.192129   & 0.1192129  & 0.4658247 &0.3383157 \\
    \hline
  \end{tabular}
   {\caption{{\bf Table 2} Migration times used in some of the simulations are indicated.
   The quantity tabulated for each planet is proportional to the product of the migration time and the mean motion     and this is held fixed.
   These values are obtained from the analytic model through an appropriate application of equations (\ref{migpres}) - (\ref{migpres2}), given an independent
   specification for the innermost planet (see Sections \ref{Migrationtimes} and \ref{Planetmasses} ). 
   The first column $(d)$  is for simulation  $sacd,$ the second $(f)$  for simulation $s_1ac_1f,$ the third $(h)$  for
   simulation  $sach.$ The final column $(m)$  is for simulations $sacm$ and $sbcm.$} }
\end{center}
\end{table}

We remark that (\ref{migpres}) - (\ref{eqntid67}) involve the mean square orbital eccentricities.
These can be related to the deviations from resonance which are in turn related to the
semi-major axes. Then these equations connect migration times to circularization times, planet masses and semi-major axes.
\subsection{Relationship between the mean square eccentricities and the deviation from resonance}\label{eccreln}
 The eccentricities may be related to deviations from resonance  using
(\ref{eqntid10})- (\ref{eqntid13}). When formulating  this we neglect variations
of the semi-major axes or equivalently only take account of variation of $x_j$ and $y_j$
assuming that the  latter  do not show any secular change.  After  time averaging 
(\ref{eqntid10}) and (\ref{eqntid11}) and making use of  (\ref{eqntid501})  we   obtain
\begin{align}
&\hspace{-0.5cm}  \left( (p_j+1)n_{j+1}-p_jn_j\right)\langle y_j\rangle 
= -\frac{\langle x_j\rangle}{t_{e,j}} \hspace{2mm} {\rm and}\label{eqntid10a}\\
&\hspace{-0.5cm}  \left( (p_j+1)n_{j+1}-p_jn_j\right)\langle x_j\rangle= 
-n_j\frac{m_{j+1}a_{j}}{M a_{j+1}}D_{j,j+1}  +\frac{\langle y_j\rangle}{t_{e,j}}.  
 \label{eqntid11a}
\end{align}
Using the above equations we obtain after straightforward algebra that
\begin{align}
&\hspace{-0.5cm}\langle e_j^2\rangle =\frac{ \left(n_jm_{j+1}a_{j}D_{j,j+1}\right)^2}
{\left(M a_{j+1}\right)^2\left(\left((p_j+1)n_{j+1}-p_jn_j\right)^2+ (t_{e,j})^{-2}\right)}\label{EJ}\\  
&\hspace{-0.5cm}{\rm for {\hspace{2mm} }}j = 1,2,3, ..., N-J.\nonumber
\end{align}
Similarly from equations (\ref{eqntid12}) and  (\ref{eqntid13})  with the help of  (\ref{eqntid501N})  we obtain
\begin{align}
&\hspace{-0.5cm}\langle e_{N-j+1}^2\rangle =\frac{ \left(n_{N-j+1}m_{N-j}C_{N-j,N-j+1}\right)^2}
{M^2\left(\left((p_{N-j}+1)n_{N-j+1}-p_{N-j}n_{N-j}\right)^2+ (t_{e,N-j+1})^{-2}\right)}\label{EN}\\  
&\hspace{-0.5cm}{\rm for {\hspace{2mm} }}  j = 1,2,3, ..., J.\nonumber
\end{align}
The mean square eccentricities  determined  from equations (\ref{EJ}) and (\ref{EN}) 
can be used in equations (\ref{migpres}) - (\ref{eqntid67}) enabling the migration times to be determined
in terms of the circularization times, masses and semi-major axes.


\begin{table}
\begin{center}
     \begin{tabular}{| c |c | c | c |c|  }
    \hline
    Migration time $\times 3n_j$ for planet $j$ & $d_1$ & $f_1$ & $d_2 $\\ \hline
   $3n_1 t_{mig,1}/10^9$  & 3.237035
              & 0.3251246
          &3.218582  \\ \hline
   $3 n_2t_{mig,2}/10^9$    & 2.561309  & 0.2569313
             & 2.569313  \\ \hline
     $ 3n_3t_{mig,3}/10^9$& 3.287614  & 0.3214347  & 3.214347  \\ \hline
     $3 n_4t_{mig,4}/10^9$& 1.315853  &0.1313746  & 1.313746  \\ \hline
     $ 3n_5t_{mig,5}/10^9$ & 1.175140
              & 0.1192129  &1.192129   \\
    \hline
  \end{tabular}
   {\caption{{\bf Table 3} Migration times,  in units of orbital period $/(6\pi)$  used in some of the simulations are indicated  (see also Table  2).    The values in the first column  $(d_1)$ are used  for simulation  $sacd_1.$
    the second $(f_1)$  is used for simulation $s_1ac_1f_1$  and  the third  $(d_2)$ for
    simulation  $sb_1cd_2.$} }
\end{center}
\end{table}


\begin{figure}
\centering
\includegraphics[trim=0.1cm 0.25cm 0.5cm 0.5cm,clip=true, height=4in,angle=0]{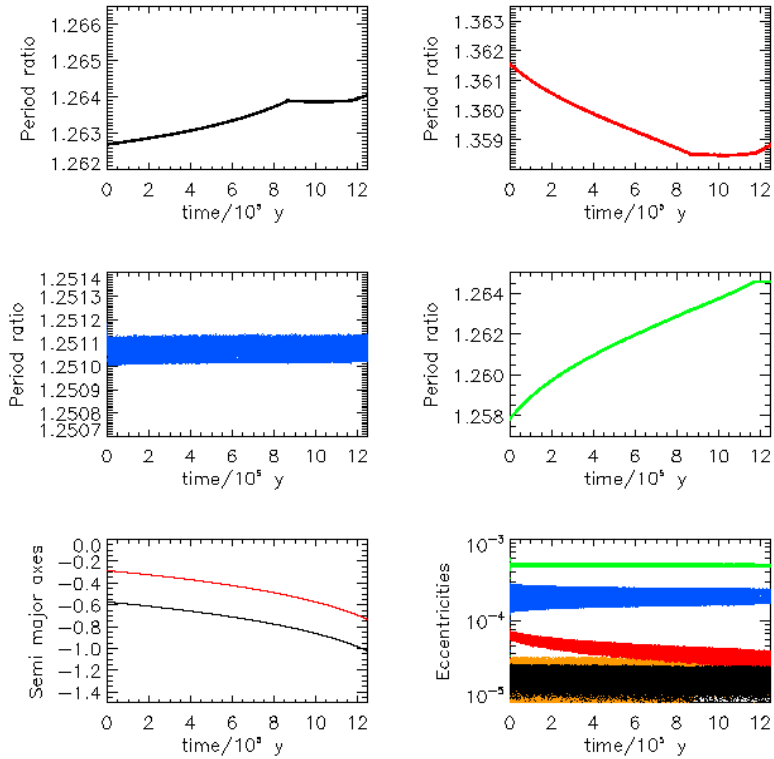}
\caption{\label{Fig1}Time dependent  evolution for run $sacd.$ The upper left   panel shows  the evolution of the ratio of the  periods of
  planets 2  and  planet 1  as a function of time. The corresponding plot for planet 3 and planet 2 is shown in the upper right panel, that
 for planet 4 and planet 3 is shown in the centre  left panel and that for planet 5 and planet 4 in the centre right  panel.
  The lower left panel shows the evolution of the logarithm of the semi-major axes of the innermost and outermost  planets in  units of $0.15 AU$
  as a function of time. The lower right panel shows the  evolution of the eccentricities   of planet 1 (black ),  planet 2 (orange ) ,  planet 3 (green),
   planet 4 (blue curve) and   planet 5 (red curve).}
\end{figure}

\begin{figure}
\centering
\includegraphics[trim= 0cm 0cm 0cm 6cm, clip=true,height=5in,angle=0]{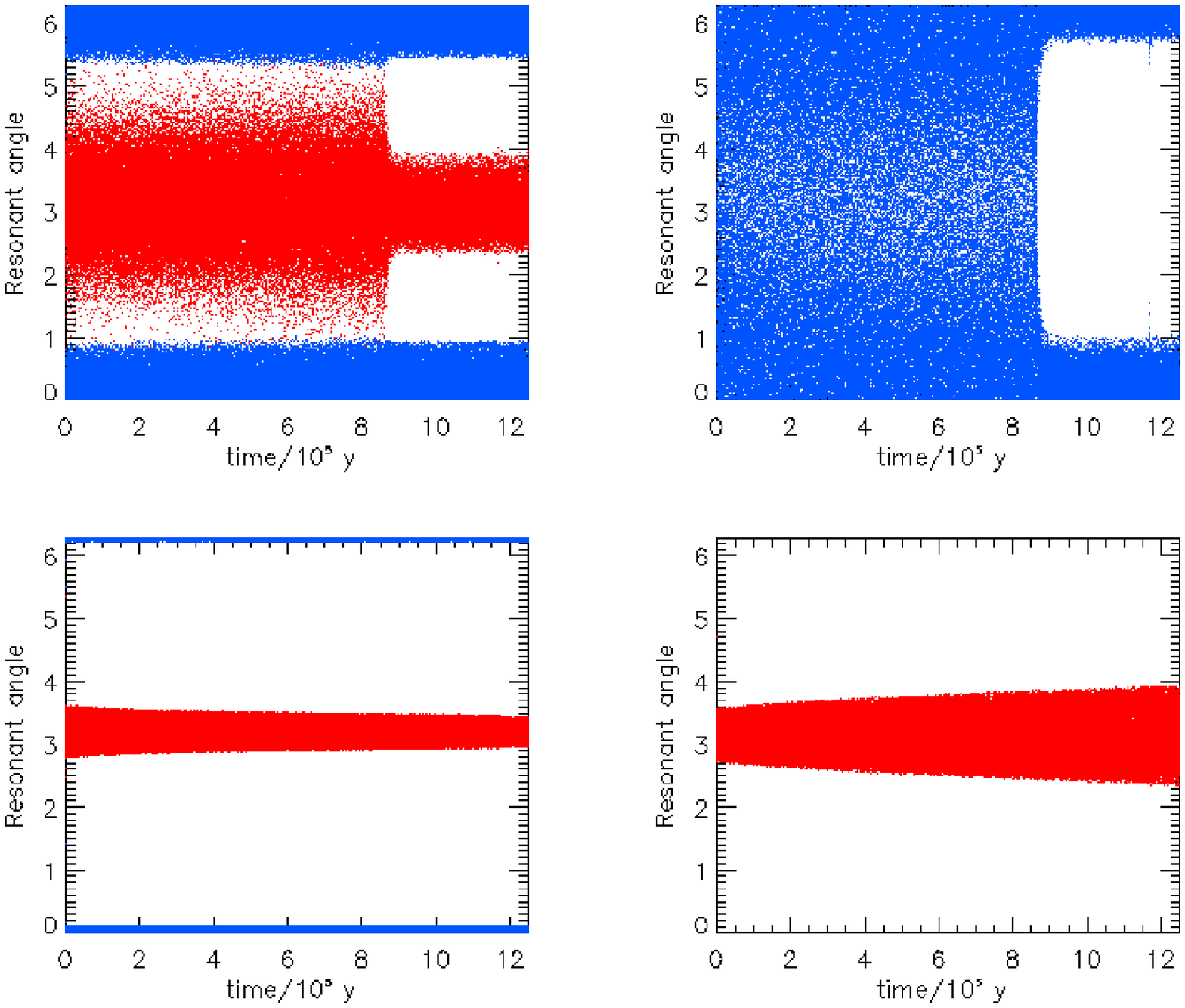}
\caption{\label{Fig2}Time dependent  evolution for run $sacd.$ The upper left panel shows the
evolution of the resonant angles $5\lambda_2-4\lambda_1-\varpi_1 $ (blue) and  $5\lambda_2-4\lambda_1-\varpi_2 $ (red). The upper right panel shows the
evolution of the resonant angle $4\lambda_3-3\lambda_2-\varpi_2 $ (blue).  The lower left panel shows the
evolution of the resonant angles $5\lambda_4-4\lambda_3-\varpi_3 $ (blue) and  $5\lambda_4-4\lambda_3-\varpi_4 $ (red).
The lower right panel shows the
evolution of the resonant angle $5\lambda_5-4 \lambda_4-\varpi_5 $  (red).
}
\end{figure}

  \section{Parameters of the simulations and application of the analytic model}\label{Simparam}
We now indicate the  setups  and parameters associated with the  models  for which simulations  were performed  and the analytic 
theory was applied.   The simulations were by means of N body calculations following the method outlined in eg. Papaloizou \& Terquem (2001)
(see also Papaloizou 2011, 2015). In all cases the system was assumed to be coplanar.
      \subsection{Initial semi-major axes and eccentricities}\label{semiaxes}
      \noindent  For the runs $sacd,$  denoted the standard case, $ sacd_1,$  $ s_1ac_1f,$  $ s_1ac_1f_1,$   $sach$ and $ sacm,$  apart from the eccentricities,
        orbital data for the five planets 
      was taken   from Campante et al. (2015) for   Kepler 444.
      The period ratios of consecutive pairs moving from innermost to outermost are then  $1.2627, 1.3615, 1.2511$ and $1.2579.$ These are close
      to 5:4, 4:3, 5:4 and 5:4 resonances respectively with relative separations of  $1\%,  2\%, 0.09\%,$and $0.6\%$ respectively. The
       third and fourth planets are accordingly significantly closer to commensurability  than the others.
       This semi-major axis setup is denoted by the label $a.$  
      
       For  the run  $sbcm$  the setup $a$ was modified such that   consecutive planetary pairs,
       with the exception of  $m_3$  and $m_4$
      were moved closer to resonance, with relative separation reduced by a factor of  $10.$   
      The period ratios of consecutive pairs  were  
      adjusted to become  $1.2513, 1.3360, 1.2511$  and $1.2508.$ This was achieved by reducing the
        semi-major axes of planets 2, 3, 4 and 5  through  multiplication by  
      factors   $ 0.993953, $ $0.981516,$ $ 0.981516$  and $0.977816$  respectively. 
      The proximity to resonance becomes comparable for all pairs. 
      This semi-major axis setup  is denoted by the label $b.$
      
      For  the run $ sb_1cd_2, $ the setup $a$  was modified such that the semi-major axis of $m_1$ was  increased by a factor 
      $1.0061.$ This has the effect of contracting the system  such that  the innermost  pair of planets
      starts closer to resonance (see below),
      This semi-major axis setup is  denoted by the label $b_1.$
    
      For all runs presented here the initial orbital eccentricities were assumed to be zero. In the case of Kepler 444 
      the observations are consistent with circular orbits
      though errors are large in the case of the outermost planet   (see Van Eylen \& Albrecht 2015).
      Note too that although the set up is for a specific set of semi-major axes, a standard scaling
      can be applied to allow results to apply to a setup where all the semi-major axes are scaled up or down by the 
      same factor (see Section \ref{Generalscaling} below).
      


\begin{figure}
\centering
\includegraphics[trim=0.1cm 0.25cm 0.5cm 0.5cm,clip=true,height=4in,angle=0]{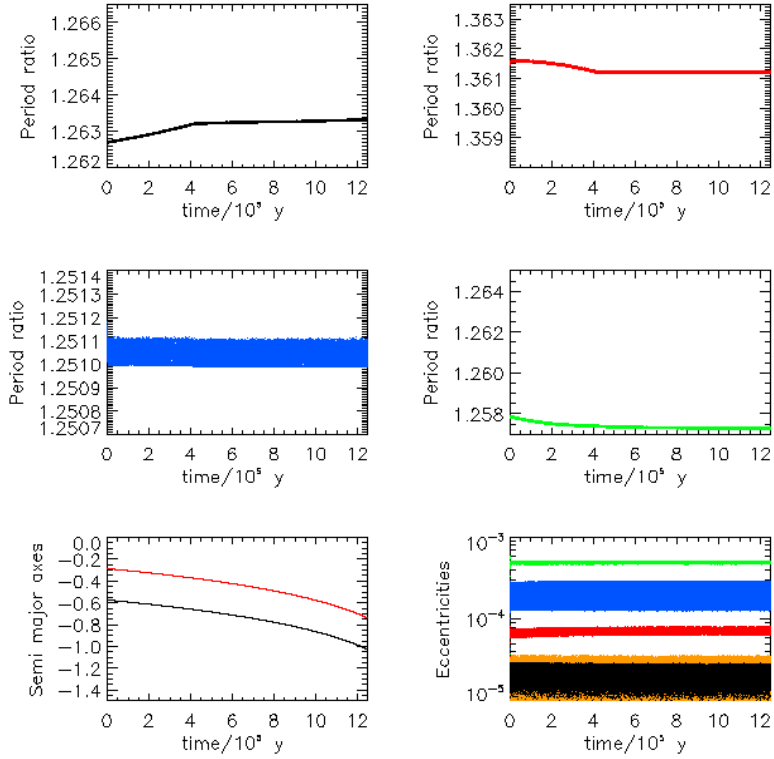}
\caption{\label{Fig3}{ As in Fig. \ref{Fig1} but for run $sacd_1.$
}  }
\end{figure}

\begin{figure}
\centering
\includegraphics[trim= 0cm 0cm 0cm 6cm, clip=true,height=5in,angle=0]{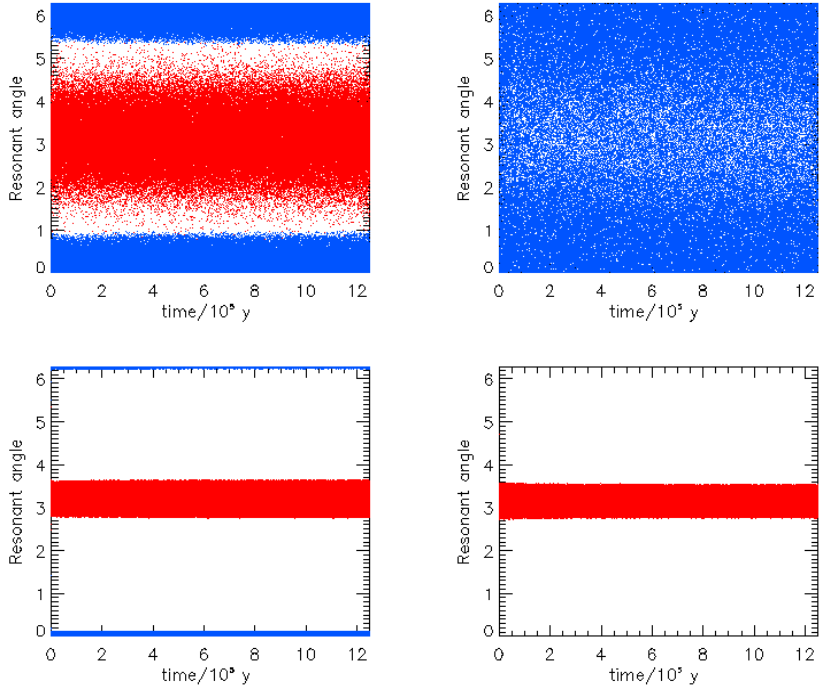}
\caption{\label{Fig4}{As in Fig. \ref{Fig2} but for run $sacd_1.$
}  }
\end{figure}


 \subsection{Circularization times}\label{circularizationtimes}
      \noindent   Following the procedure  outlined in Section \ref{circtime} we adopt the disk model 
       for which $t_{e,j}$ scales as $1/n_j$ or equivalently the orbital period of planet $j$ (see equation (\ref{circmodele})).
       Migration times obtained from the analytic model of Section \ref{sec5} will then scale in the same way.
      The standard circularization times  $(c)$  we adopted for planet, $j,$  are  given by 
        \begin{equation}  t_{e,j}=6.4\pi\times10^{-4}
       (M/m_j)/n_j.\label{circ}
       \end{equation}
       This is equivalent to using (\ref{circmodele}) adopting a disk with a mass $5M_J$ within $1AU$
       being a factor $5.6$ larger than expected from a minimum mass solar nebula. A disk this massive was adopted 
       so as to obtain characteristic evolution times for the system $\sim 10^{6-7} y.,$ as conventionally expected for a
       protoplanetary disk lifetime, when the procedure to determine migration times described in Section \ref{Migrationtimes}
       below was followed.
       Equation (\ref{circ})  
       was  used in all cases except  $s_1ac_1f$  and  $s_1ac_1f_1$ for which
       the values obtained from this formula were increased 
       by a factor of $100$ (see Table 1).

      \subsubsection{Migration times}\label{Migrationtimes}
      The migration times for the runs  $sacd,     sach,   sacm $ and $ sbcm$   indicated  in Table 1
       and listed in Table 2  were determined 
       from the analytic theory through the use  of equations (\ref{migpres}) - (\ref{migpres2}). 
       In all cases excepting $ sacm $ and $ sbcm$ all consecutive pairs  are assumed   to interact 
       resonantly  with  $p_1=p_3=p_4 =4$ and $p_2=3.$
       In addition $J= 2$ so that only planets $3$ and $4$ are coupled by two first order resonances.
       This choice was made as planets $3$ and $4$ are the closest to resonance
       in the Kepler 444 system.
       
       The cases $sacm $ and $sbcm$ were calculated assuming the system was composed of two separate  independent 
       subsystems as outlined in Section \ref{subsystems} above.
       In these cases planets $1$ and $2$ as well as planets $3$ and $4$  are coupled by a pair of first order resonances. 
       We recall that migration times 
       can be  determined once the semi-major axes, masses and circularization times are specified, 
       given a migration time for one of the planets in each subsystem,  
       here taken to be the innermost one, assuming that the period ratios remain constant.
       More about the adopted masses  and this procedure is indicated below.
       
      The migration times for the run  $s_1ac_1f$   listed in Table 2 were obtained from the set $d$ by reducing 
      them by a factor of ten. 
       This is to test the application of a simple
      scaling relation expected from the simplified analytic theory.
       The migration times for the runs $sacd_1,$  $s_1ac_1f_1$ and $ sb_1cd_2$  are
        labeled $d_1, f_1$ and $d_2$ respectively and are listed in Table 3. 
       These are obtained by slightly modifying  values obtained from the analytic theory.
       Thus  the times $d_1$ are obtained by heuristically adjusting those  obtained from $d$ so as to reduce the magnitude  of the evolution of the period ratios.  
       The times  $f_1$ are  obtained from $f$ by increasing the migration time of the innermost planet by
        a factor $1.001.$ 
       The times $d_2$ are obtained from $d$ by rescaling the migration time of the innermost planet 
       so that was the same at  its shifted position in $b_1$
       as in its original position in $a.$
      Note that the modification in the latter two cases only affects the migration time of the innermost planet.

      \subsubsection {Planet masses}\label{Planetmasses}
      As only planet radii are available and their densities are unknown,  planet masses cannot be determined.
      In order to proceed we determined a set of planet masses by trial and error that were such that the product of the masses  and  migration times
      determined by application of the analytic model of Section \ref{sec5},  through use of equations (\ref{migpres}) - (\ref{migpres2}) in Section \ref{anmig} 
      , for the standard run $sacd,$ were constant to within $\sim 3\%.$ This would be expected from the simplest migration calculations (eg. Papaloizou \& Terquem 2006).
      In carrying out this procedure  the migration time of the innermost planet, which was  not otherwise determined,  was chosen such that the system
       as a whole migrated significantly over an expected protoplanetary disk
      lifetime.   Although the analytic model,  through determination 
      of the relative migration rates  constrains the planet masses,  
      we stress that the parameters we adopt are by no means unique but they enable investigation of the orbital architecture of a system resembling Kepler 444
      and tests carried out for other examples not considered here have 
      been found to lead to very similar conclusions.

\begin{figure}
\centering
\includegraphics[trim=0.1cm 0.25cm 0.5cm 0.5cm,clip=true,height=4in,angle=0]{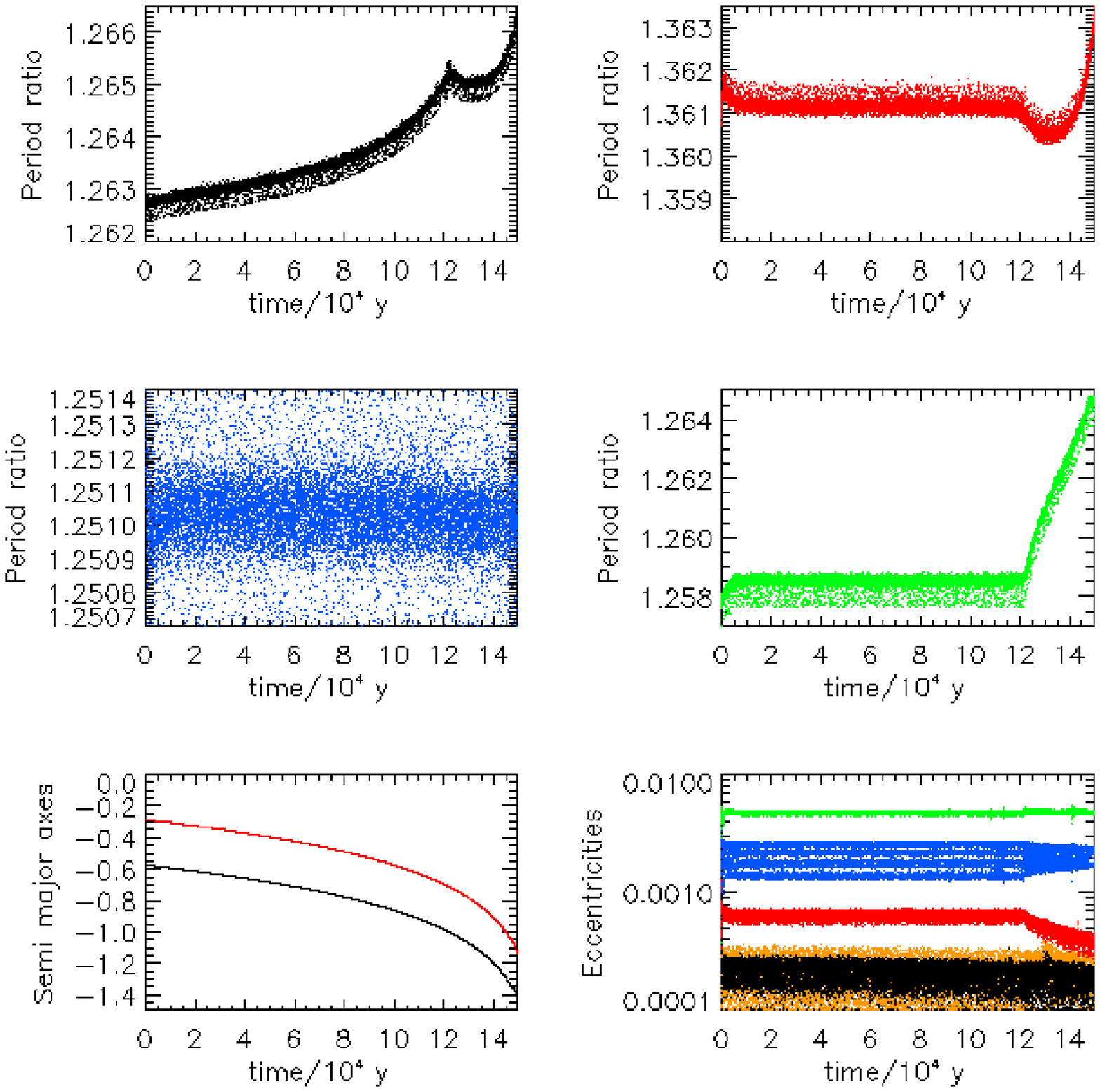}
\caption{\label{Fig5}{As in Fig. \ref{Fig1} but for run $s_1ac_1f$.  Note that the range of eccentricities 
shown is an order of magnitude larger in this case.  }}
\end{figure}

\begin{figure}
\centering
\includegraphics[trim= 0cm 0cm 0cm 6cm, clip=true,height=5in,angle=0]{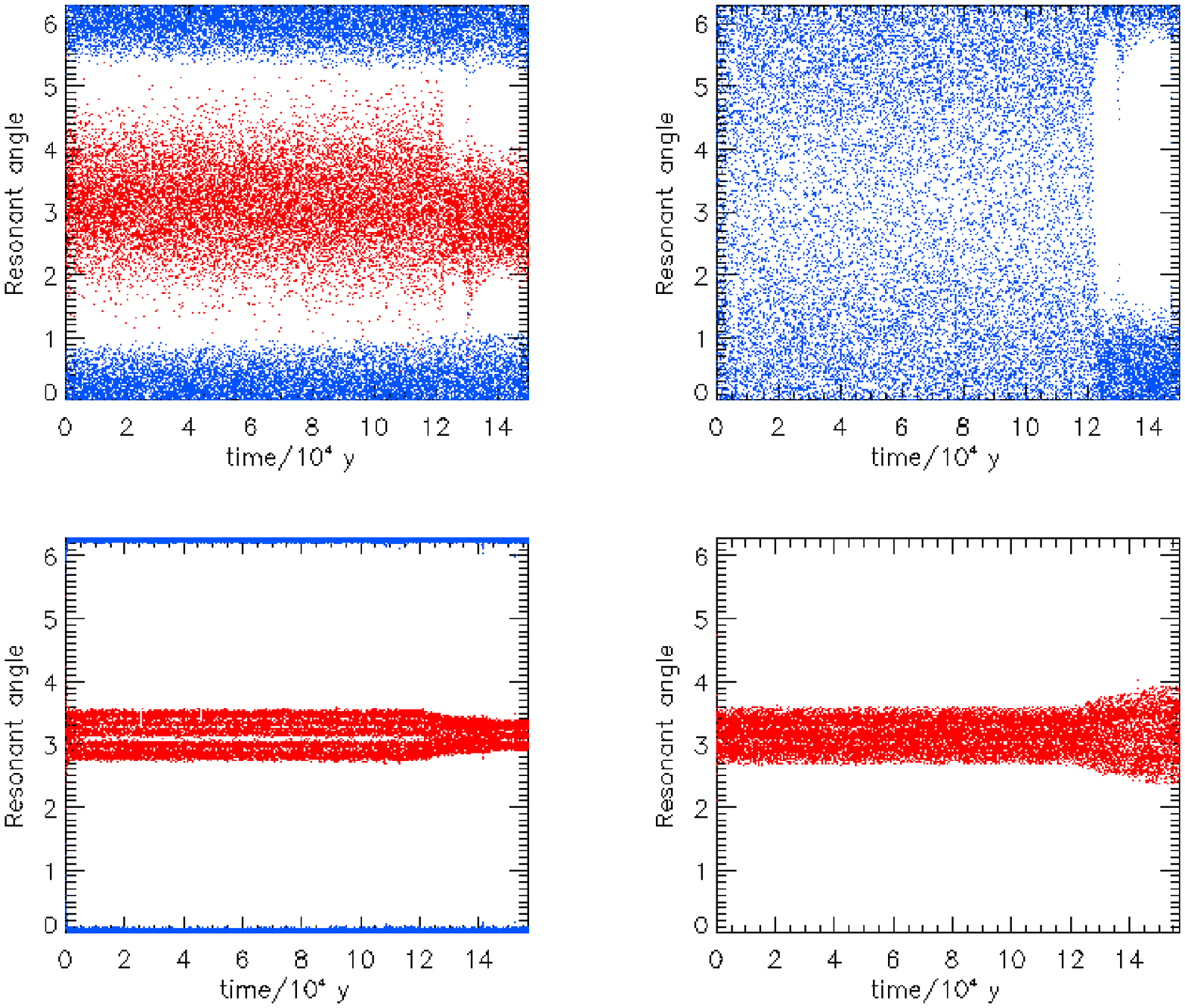}
\caption{\label{Fig6}{As in Fig. \ref{Fig2} but for run  $s_1ac_1f.$
}  }
\end{figure}

 The  standard masses    
          chosen alongside  the determination of migration times,  using the analytic procedure described in Section (\ref{anmig}),
 as indicated above, being  denoted by the label $s,$ were
 \noindent   $ m_1~=~0.0546075M_{\oplus},$
     $ m_2~=~0.0689779M_{\oplus},$
      $ m_3~=~0.05333708M_{\oplus},$
     $ m_4~=~0.1348677M_{\oplus},$ and  
     $ m_5~=~0.1479524M_{\oplus}.$
      \noindent    Adopting the planetary radii given by Campante et al. (2015),  the mean densities of the planets  are then
         found to be $\rho_j/\rho_{\oplus} = 0.853,  0.562,  0.369,   0.829$ and  $ 0.364 $
        for $j = 1,2,3,4$ and $5$ respectively,   indicating that they could be rocky bodies.  
         We remark that these densities  differ by a factor $\sim2$ varying non monotonically  with orbital separation.
        This type of non uniformity   has been  observed in  the  Kepler  11 system  (Lissauer et al. 2011b) and the Kepler 138 system (Jontof-Hutter et al. 2015)
        and so should not be unexpected. 
        Standard masses  were adopted in all cases  considered here except for $s_1ac_1f$  and $s_1ac_1f_1$ for which they were increased by
      by a factor of $10$ (see  Table 1).
      
      \subsection{General scaling}\label{Generalscaling} 
      Although the setups described above are for specific sets of semi-major axes,
      because the circularization and migration times we adopt are proportional to the orbital period,
      a standard scaling can be applied to expand the length scale of the initial setup.
      This simple scaling holds masses fixed while length scales are increased by a factor $\Lambda $ 
      and times are increased by a factor $ \Lambda ^{3/2}.$
      Thus results obtained for a given setup can be easily scaled to apply to an initial setup with
      an expanded length scale. This can also be arranged so that a given simulation is scaled to terminate
      close to the original setup.

\begin{figure}
\centering
\includegraphics[trim=0.1cm 0.25cm 0.5cm 0.5cm,clip=true,height=4in,angle=0]{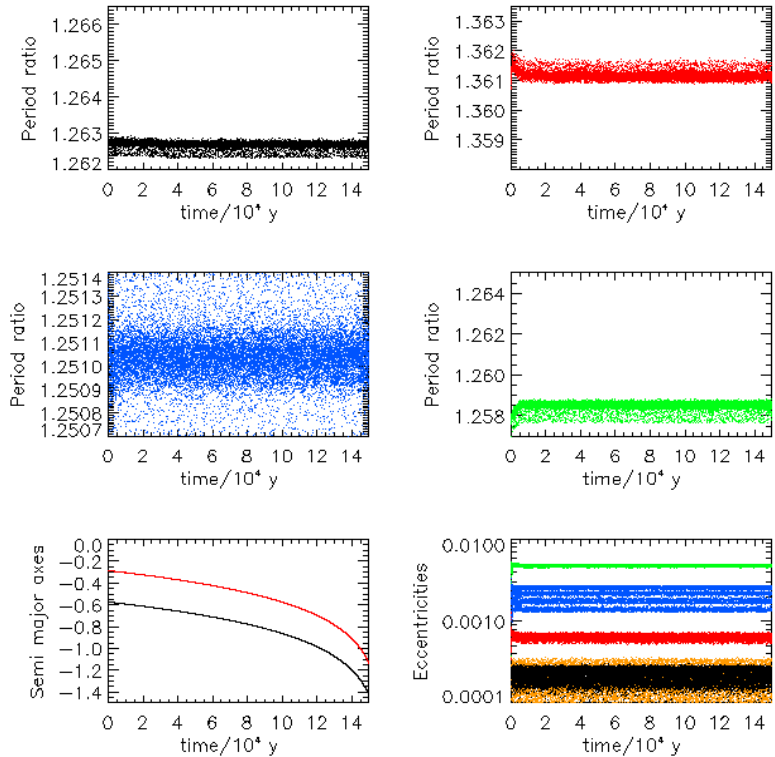}
\caption{\label{Fig7}{As in Fig. \ref{Fig1} but for run    $s_1ac_1f_1.$ Note that the range of eccentricities 
shown is an order of magnitude larger in this case. 
}  }
\end{figure}

\section{Simulation results}\label{simres}
We now  describe  results for the standard case $sacd.$
\subsection{The standard case $sacd$ }\label{stan}
For this case the initial semi-major axes and eccentricities  were set up as indicated in Section \ref{semiaxes} (see Table 1) 
and circularization times as indicated in Section \ref{circularizationtimes}.
Migration times were determined as indicated in Section \ref{Migrationtimes}  and are given in Table 2. The planet masses are given in Section \ref{Planetmasses}.
We remark that the migration time of the innermost planet given in Table 2 is such that  $t_{mig,1}/t_{e,1} =   1.3\times10^5 .$
The expected value from the simplest migration calculations  described in Sections  \ref{sec3}  and \ref{circtime} 
for which corotation torques do not play a significant role is $\sim (R/H)^2 \sim 2.5\times 10^3 $ which is smaller by a 
factor $\sim 54.$  This is an  indication that 
modifications  along the lines described
in Section \ref{sec3}   may need to be invoked in order
 to  slow down inward migration by between one and two orders of magnitude. 


The time dependent evolution of the semi-major axes of the innermost and outermost planet is plotted in Fig. \ref{Fig1}.
The overall contraction of the system is by a factor of $\sim 3$  for a run time of $1.25\times 10^6 y.$ 
Note that the system can be scaled to start at initial radii a factor of two larger by increasing 
the times by a factor $2^{3/2},$
then finishing near to the initial set up.
The  time dependent evolution of ratio of the  period ratios of
 planet 2  and  planet 1,  planet 3 and planet 2,  as well as  
 planet 4 and planet 5 are also  shown  in Fig. \ref{Fig1} as is the time dependent evolution of the eccentricities. 
   
The eccentricities obtained from the analytic model were
 $e_1 = 2.01\times 10^{-5}, e_2 = 7.41\times10^{-6}, e_3 = 4.81\times 10^{-4}, e_4 =   2.04\times 10^{-4},$ and $e_5 =   6.70\times 10^{-5}.$
 These are in reasonable agreement with the mean  values seen at early times for simulation $sacd$ and at all times for simulation $sabcd_1$  described below
 for which mean values of the  eccentricities show less variation. These small values are  also in line with the near circular orbits inferred from the observations by
 Van Eylen \& Albrecht (2015).

The quantity  ${\cal C}_{N,1}  = (1/t_{mig,N}- 1/t_{mig,1})t_{mig,1}$ is a measure of the rate of convergence of the entire system. For the run $sabc,$ 
${\cal C}_{N,1} =   7\times10^{-3}$ (see Table 2). This means that a scale contraction of order unity is associated with a relative contraction
of only around $1\%.$    This is a consequence of the relatively wide separation of consecutive pairs from resonance.
From equation  (\ref{eqntid66N1}) it follows that ${\cal C}_{N,1}$  scales as the squares of the eccentricities which themselves are  
approximately inversely proportional to the separation from resonance (see equations (\ref{EJ}) and (\ref{EN}) ).
Thus in order to obtain a faster rate of convergence of the system, enabling significant convergence  during the evolution, 
from our analysis the system would have had to have been closer to resonance. Only in the late stages of evolution could the migration times
increase to produce conditions like those of $sacd.$

The time dependent evolution of the resonant angles $5\lambda_2-4\lambda_1-\varpi_1 ,$   $5\lambda_2-4\lambda_1-\varpi_2 ,$ 
 $4\lambda_3-3\lambda_2-\varpi_2, $  
 $5\lambda_4-4\lambda_3-\varpi_3 ,$  $5\lambda_4-4\lambda_3-\varpi_4 $ 
and  $5\lambda_5-4 \lambda_4-\varpi_5 $  is shown  in Fig. \ref{Fig2}.
All of these angles,  except the second  connecting planets $1$  and $2,$  contribute in the analytic theory and apart from that one, they are all seen to librate at early times
except 
 for $4\lambda_3-3\lambda_2-\varpi_2 .$
This angle circulates.  In addition  although the angle,  $5\lambda_2-4\lambda_1-\varpi_2, $
 circulates   at early times  as expected   according to the analytic model of Section \ref{sec5} 
  (see  discussion in Section \ref{sec4.3} below equation (\ref{LaLa}) 
and also in Sections \ref{subsystems} and \ref{Migrationtimes}), it does not do so uniformly indicating that it may contribute to the evolution. 
 The last two features indicate  that  at this stage,  planets $1$ and $2$ are
 not as strongly coupled to the rest as in the simple analytic theory,   a feature further  investigated below. 
 This may be related to the fact that planets $2$ and $3$ are the most widely
 separated from resonance.  Resonant angles not illustrated in Fig. \ref{Fig2} always circulate.
 
  The deviation of the period ratio of planets $4$ and $5$ from strict resonance increases by a factor $\sim 1.8$ during the run.
 Corresponding to this change  the  mean eccentricity  of planet $5$ decreases by  approximately the same factor  as expected from
 equation (\ref{EN})  when  the circularization time  there is assumed to be arbitrarily large. 
 The  deviation of the period ratio of planets  $1$ and $2$ from resonance increases by $ \sim10^{-3}$
 corresponding to a  relative increase of $\sim 8\%.$ The mean eccentricity of planet $1$ is seen to decrease by a corresponding amount  as indicated by
 equation (\ref{EJ}).
  The deviation of the period ratio  of planets $2$ and $3$ from resonance
 decreases by $ \sim 2\times 10^{-3},$ over the run time.   However, this deviation as well as that  associated with planets $1$ and $2$
 stop  their secular advance  at a time  $\sim 8.5\times 10^5 y.,$ 
 at which point a Laplace resonance between planets $1,$  $2$ and $3$ is formed. This is indicated by the fact that after that 
 the angles  $4\lambda_3-3\lambda_2-\varpi_2 $ and  $5\lambda_2-4\lambda_1-\varpi_2 $ both then clearly librate.
 The period ratio of planets $3$ and $4$ being the closest to resonance
 remains essentially unchanging throughout the evolution.  The period ratio of planets $4$ and $5$ increases throughout,   
 stabilizing only at the end of the simulation, the  deviation from resonance  having  increased by  $\sim 5\times 10^{-3}.$
 The setting up of a Laplace resonance was not anticipated in the analytic treatment. However, we found that
small  adjustments to the input migration times could result in much smaller changes  to the period ratios,
so avoiding the formation of a Laplace resonance.  
\begin{figure}
\centering
\includegraphics[trim= 0cm 0cm 0cm 6cm, clip=true,height=5in,angle=0]{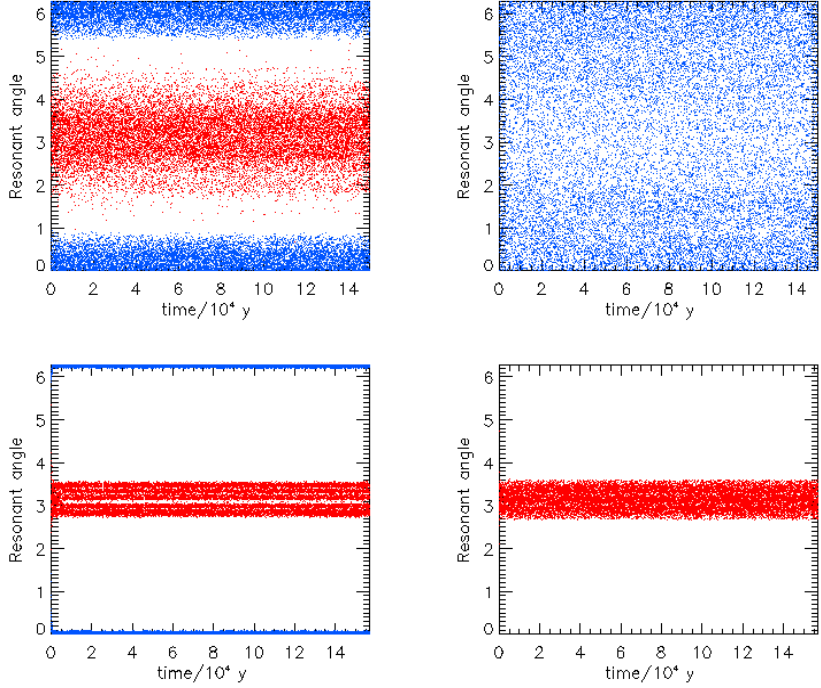}
\caption{\label{Fig8}{As in Fig. \ref{Fig2} but for run $s_1ac_1f_1$.
}  }
\end{figure}

\subsubsection{The modified standard case $sacd_1$}\label{modstan}
In this case the initial setup and initial  conditions  were the same as for $sacd,$
the only difference was that the input migration times were slightly adjusted (see Section \ref{Migrationtimes}).
The modified values are listed in Table 3.
 In this case the simulation underwent the same  amount of radial contraction
but the period ratios underwent  much less evolution. The maximum  deviation from resonance changing by  less than  $5\times 10^{-4}$
in all cases. These features are illustrated in Fig. \ref{Fig3}. The evolution of the resonant angles 
illustrated in Fig. \ref{Fig4} is similar  to that found for  $sacd$ 
at early times. But in this case the variation of the period ratios is insufficient 
for a Laplace resonance to  form and so the character of the evolution of the resonant angles does not change.

Although these simulations indicate that it is possible to set up the system such that convergent migration
and circularization are balanced. The wide separations from resonance characteristic of Kepler 444
indicate only very weak convergence. This means that if the system is not  formed very close to it's final period ratio
configuration,  the migration times assumed for $sacd_1$ could only apply in the final  evolutionary stages.    

\begin{figure}
\centering
\includegraphics[trim=0.1cm 0.25cm 0.5cm 0.5cm,clip=true,height=4in,angle=0]{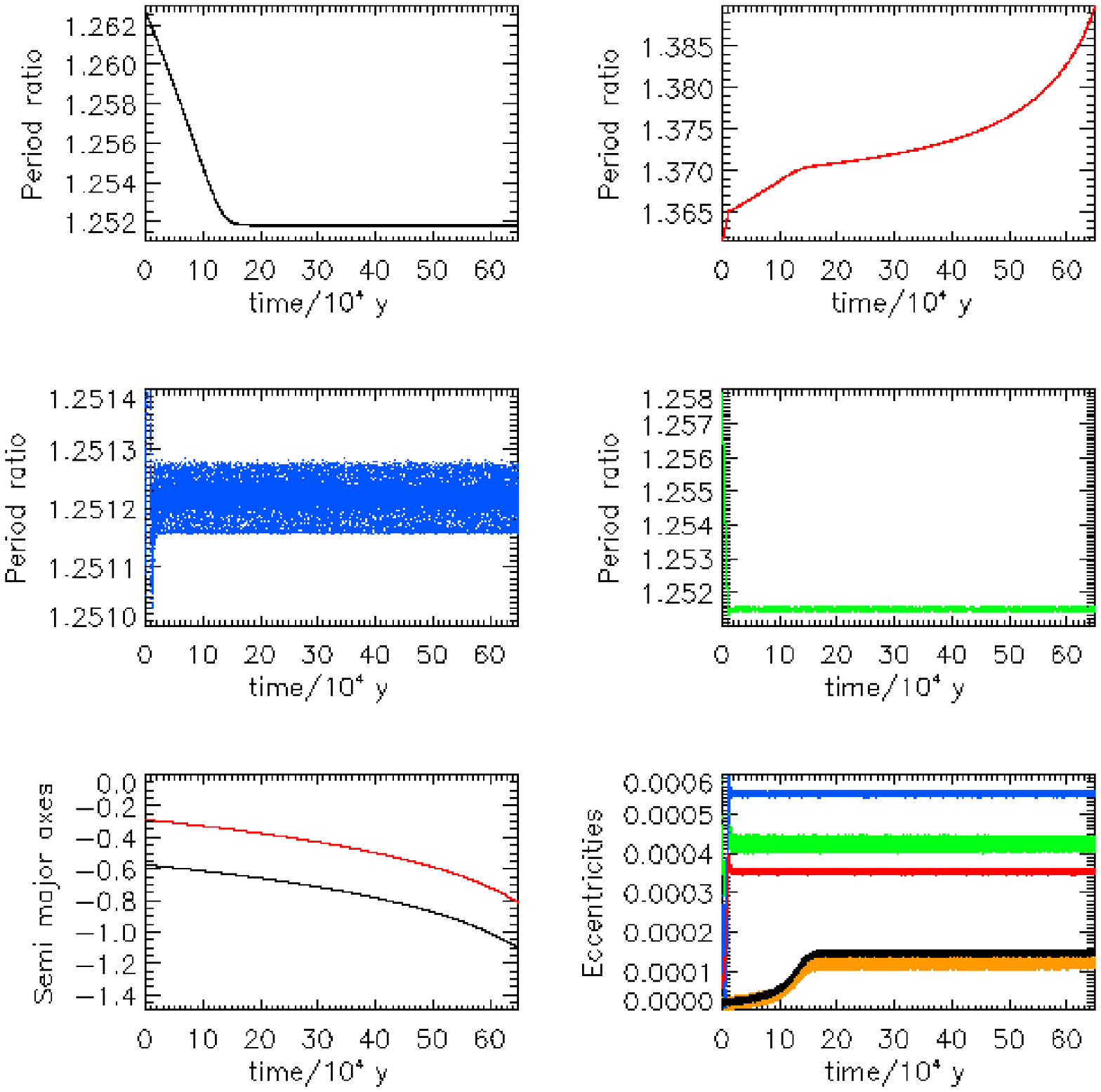}
\caption{\label{Fig9}{As in Fig. \ref{Fig1} but for run $sach.$
}  }
\end{figure}

\subsection{Simulations with larger masses}\label{lmsims}

According to the analytic discussion in Sections \ref{sec5} and \ref{anmig}, when  the  planet masses are 
increased  by a factor, $F_m, $
and the expression for the circularization time given by equation (\ref{circ}) is  scaled by a factor $F_c,$ 
the eccentricities scale as $F_m$ as long as the  inverse circularization time in the denominators of equations (\ref{EJ}) and (\ref{EN}) is neglected,
the latter being a good approximation. 
 The calculated convergent migration rates then scale as $F_m^3/F_c.$ Thus if all masses are scaled by a factor of $10$ and $F_c=100,$
calculated  migration times should consistently decrease by a factor of  $10.$   

 In order to test this scaling  we performed run  $s_1ac_1f$ for which the input data for run $sacd$  was scaled as indicated above.
 Planet masses were increased by a factor of $10,$ migration times were reduced by a factor of $10$ and $F_c=100$ see Tables 1 and 2.  

The time dependent evolution of the semi-major axes of the innermost and outermost planet and the period ratios is illustrated in  Fig. \ref{Fig5}.
The overall contraction of the system is by a factor of $\sim 3$   after a time  $1.25\times 10^5 y.$ consistent with
the migration time reduction expected  when compared to run $sacd.$
Note that as for that case,  the system can be scaled to start at larger  initial radii  by appropriately increasing the times
so  that the system then finishes near to the initial set up.
In this case, with the exception of the innermost pair, the period ratios of consecutive pairs  on average remain constant
up and till $1.2\times 10^5 y.$  This  indicates that the innermost pair are more strongly coupled than in the analytic model
which appears to work better for the other planet pairs in this case.
 Up till this time the innermost period ratio increases  steadily until the deviation from resonance is  $\sim 2\times 10^{-3}.$
Then, once the conditions for it are satisfied,  a Laplace resonance is set up as for the run $sacd.$
After this, the period ratio deviation from resonance of planets $2$ and $3$ and also  $4$ and $5$ increase, the latter by up to  $5\times 10^{-3}.$
The time dependent evolution of the resonant angles corresponding to those for the run $sacd$ shown in Fig.\ref{Fig2} 
is shown  in Fig. \ref{Fig6}. The evolution of all angles is qualitatively very similar in  these  two cases.
 However, the  Laplace resonance is  set up
after  $1.2\times 10^5 y.$ which is somewhat later than the expected scaled time of $8 \times 10^4 y.$ 
When this occurs the angle $4\lambda_3 - 3\lambda_2 - \varpi_2$ starts to librate causing the form of the evolution to change.
Planets $2$ and $3$ become more strongly coupled , affecting  planet $4$ through the strong resonance between  planets $3$ and $4,$
and hence the period  ratio of planets $4$ and $5.$  The overall effect of the increased resonant coupling  is that the system separates.

\subsubsection{ Reducing the migration rate of the innermost planet}\label{redin}

The Laplace resonance was set up in the run $s_1ac_1f$  because of the evolving period ratio of the innermost pair of planets.
In order to counteract this and obtain more stable period ratios we repeated the simulation with the migration time of the inner planet increased by a factor
$1.001$ in simulation $s_1ac_1f_1.$  The evolution of the  period ratios  plotted in Fig. \ref{Fig7} shows that the   period ratio deviations from resonance
change by less than $\sim 5\times 10^{-4}$ in all cases with no Laplace resonance being set up.
The time dependent evolution of the  resonant angles corresponding to those shown for the run $sacd_1$ is shown in Fig.\ref{Fig8}, 
the character of the evolution being very similar.
Although the simulations for larger planet masses show similar features to those with lower masses, there is not a precise scaling
between the two most likely because the simple analytic model does not give a completely accurate 
description for the behaviour of the resonant angles.

\subsection{Simulations with closer commensurabilities}\label{closerr}
\begin{figure}
\centering
\includegraphics[trim= 0cm 0cm 0cm 6cm, clip=true,trim=0.1cm 0.25cm 0.5cm 0.5cm,clip=true,height=4in,angle=0]{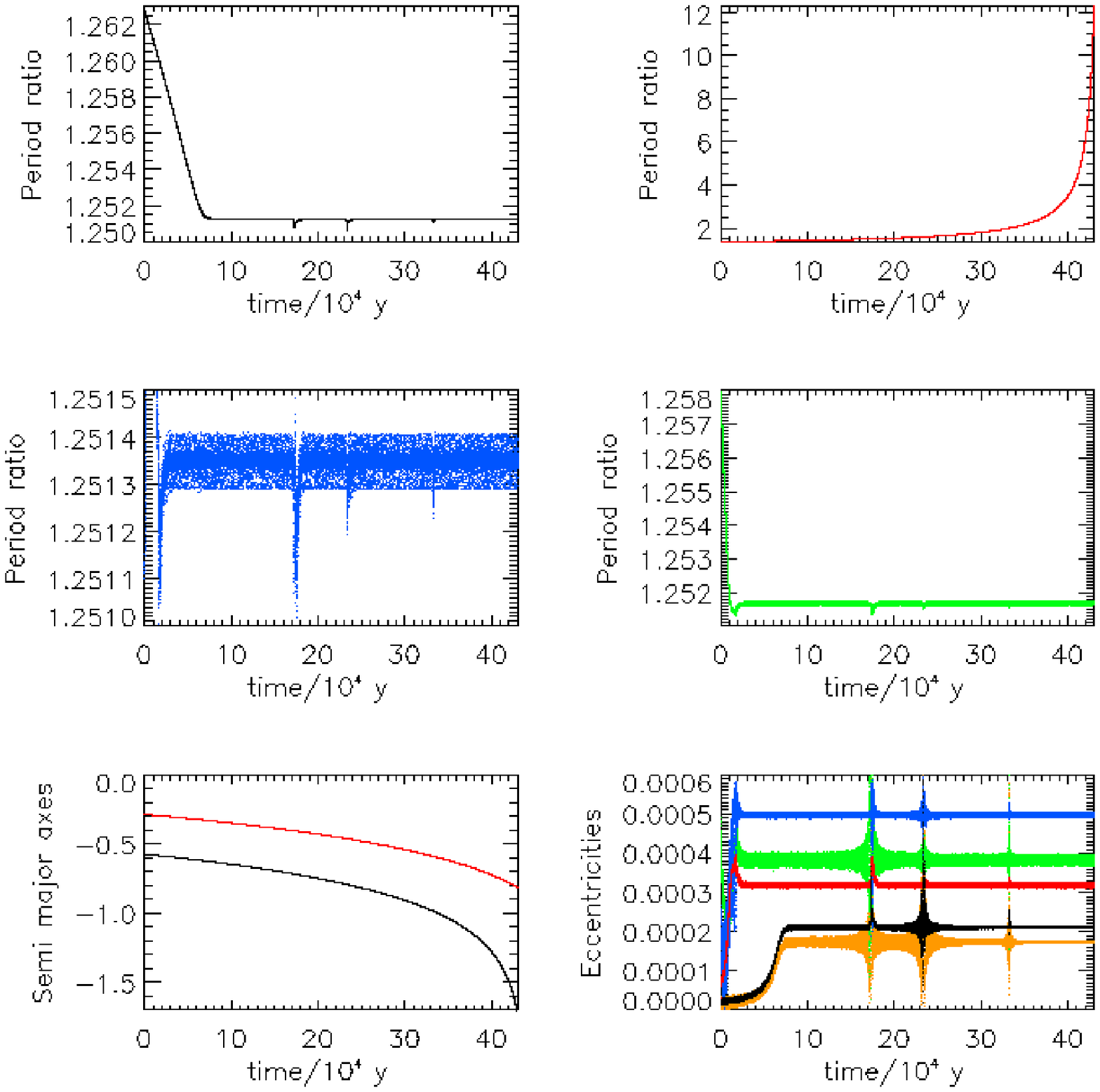}
\caption{\label{Fig13}{As in Fig. \ref{Fig1} but for run $sacm.$
}  }
\end{figure}
Although we have been able to obtain migrating systems with only a small variation of period ratios,
on account of the  deviations from resonances being too large, the inferred  rates   of convergent
migration are  very small implying the system cannot have been brought together from a wide separation.
Such a situation could only apply  during the late evolutionary stages. In order to study systems with 
faster relative migration rates we have performed simulations  $sach,$  $sacm$ and $sbcm$
which have  systems with consecutive pairs,  except for planets $3$ and $4,$  an order of magnitude 
closer to resonance than those considered above.  

In order to obtain migration times from the analytic method
of Section \ref{sec5} for these, the initial semi-major axes of consecutive pairs  were
adjusted so that the  deviations of all consecutive pairs of planets from resonance became $\le 2\times 10^{-3}$
as indicated in Section \ref{semiaxes} for  the procedure labelled  $b.$ 
For these simulations, for simplicity   we retain the same planet masses as in the standard case even though the dependence on migration time
is changed. We note that this could be adjusted by adjusting the relative separations from resonance for the different planets.

\begin{figure}
\centering
\includegraphics[trim=0.1cm 0.25cm 0.5cm 0.5cm,clip=true,height=4in,angle=0]{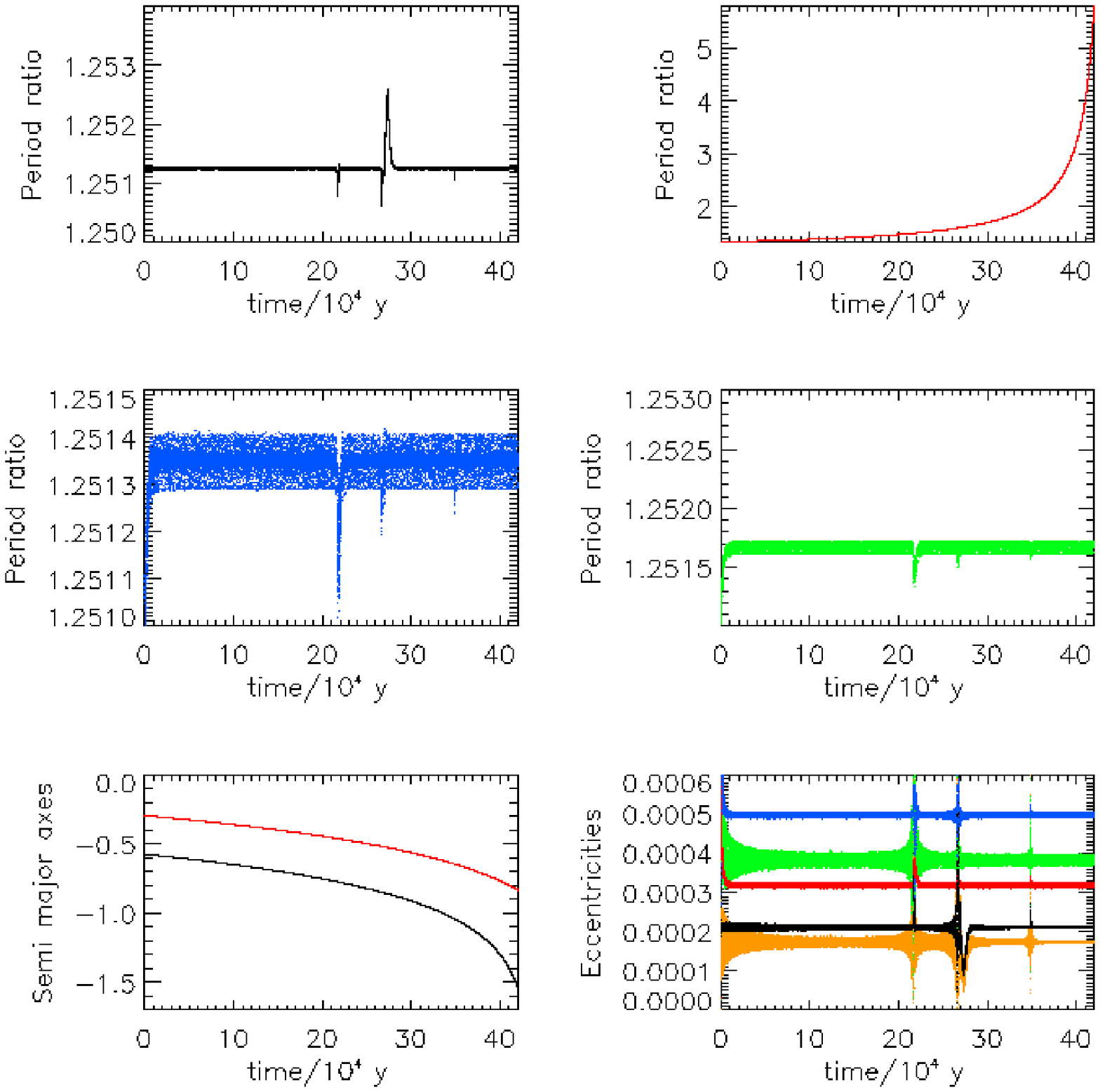}
\caption{\label{Fig15}{As in Fig. \ref{Fig1} but for run $sbcm.$ }  }
\end{figure}
\subsubsection{Simulations with migration rates obtained assuming all consecutive pairs  are coupled}\label{allcoupled}
The migration times obtained following the procedure for this case  outlined in  Section \ref{subsystems}
labelled, $h,$  are tabulated in Table 1.
However, when  performing the simulation $sach,$  , the semi-major axes were initiated as in procedure $a$ (see Section \ref{semiaxes}).
 The consequence of that  in this case  is that the period ratios adjust to become closer to resonance.
 The relative convergence parameter ${\cal C}_{N,1}  = 0.3 $ indicating that a significant relative contraction
 can occur while the system contracts as a whole.  
The time dependent evolution of the semi-major axes of the innermost and outermost planet and the period ratios is 
plotted in Fig. \ref{Fig9}.
The overall contraction of the system is by a factor of $\sim 3$   after a time  $6.5\times 10^5 y.$ 
In this case the period ratios for the innermost and outermost pairs of planets  move rapidly closer to resonance as expected.
However, the period ratio of planets $2$ and $3$ continually increases   up  to $1.39$ indicating that the inner two planets decouple from the outer $3$
as indicated above. This is supported by the behaviour of the resonant angles which all librate except for those connecting planets $2$ and $3.$
Accordingly we go on to consider simulations for which the migration times are calculated assuming this decoupling occurs.

\subsubsection{Migration rates obtained assuming inner two and outer three planets behave as separate systems}\label{sepsubs}
The migration times obtained following the procedure for this case  outlined in  Section \ref{sec5}
are labelled, $m$ and are tabulated in Table 1. As indicated in  Section \ref{subsystems} in order to obtain these,
the migration times for both planet 1 and planet 3 should be specified.  This determines
how the separate subsystems separate. To obtain a working model we specified $t_{mig,3}=0.8t_{mig,1}.$ 
When  performing the simulation $sacm,$   the semi-major axes were initiated as in procedure $a.$
On the other hand, when  performing the simulation $sbcm,$ the semi-major axes were initiated as in procedure $b$
as used for the analytic calculation of the migration times.  The relative convergence parameter ${\cal C}_{N,1}  = 0.35 $
for the innermost pair which is significant.
The time dependent evolution of the semi-major axes of the innermost and outermost planet and the period ratios  for $sacm$ is plotted in  Fig. \ref{Fig13}.
The corresponding quantities for $sbcm$ are plotted in Fig. \ref{Fig15}.  Apart from planets $2$ and $3$ the period ratios move closer to resonance for $sacm$ but remain
approximately constant for $sbcm.$ The period ratio of planets $2$ and $3$   increases dramatically up to $5$
for $sacm$  and $10$ for $sbcm$ indicating a more serious decoupling of the two subsystems  than for $sach.$ 
However, the large increase in this period  results in successive passage through 3:2, 5:3 and 2:1 resonances which are indicated by the spikes seen in 
time  dependent evolution of the period ratios.  The behaviour of the resonant angles  for both simulations is very similar to that seen in $sach.$
\begin{figure}
\centering
\includegraphics[trim=0.1cm 0.25cm 0.5cm 0.5cm,clip=true,height=4in,angle=0]{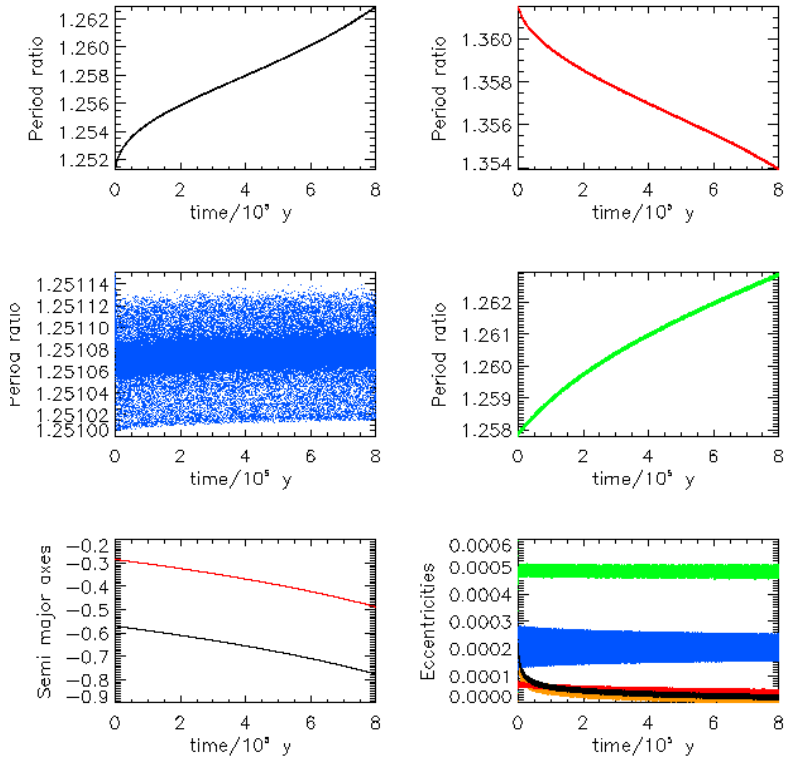}
\caption{\label{Fig17}{ As in Fig. \ref{Fig1} but for run $sb_1cd_2.$}  }
\end{figure}
\subsubsection{ The innermost planet moved closer to resonance}\label{incloserr}
The above simulations indicate the system should have been closer to resonance than  currently observed   
in Kepler 444 if significant convergent migration has occurred.
To attain a situation like that in Kepler 444, the migration times would have to increase to attain values like those in setup $a,$
at which point little further convergence occurs. In order to investigate whether such adjustments can occur
we performed simulation $sb_1cd_2.$ This had semi-major axes set up as in $sacd$  but the  semi-major axis of the innermost planet 
then increased such that the period ratio of planets $1$ and $2$ was the same as  for set up $b.$
The migration time of the innermost planet was then scaled   such that  it took on the same value  as  in  $sacd,$ 
but at  its new position.
The time dependent evolution of the semi-major axes of the innermost and outermost planet and the period ratios  for $sb_1cd_2$  are  plotted in Fig. \ref{Fig17}. 
The period ratio for planets $1$ and $2$ increases to the current value after $8\times 10^5 y.$
However the period ratio of the outermost pair also increases significantly while the  period ratio of planets $2$ and $3$
decreases significantly. Although the above does not  represent a precise modelling, it does indicate that expansion to present period ratios
from ones closer to resonance is a possibility.

\section{Discussion}\label{discu}

In this paper we have studied the evolution of multi-planet systems in which the components
undergo  tidal interaction with a protoplanetary  disk with reference to a system 
with architecture resembling that of Kepler 444.

We formulated  an analytic model for a planetary system with consecutive pairs in resonance
undergoing orbital  circularization and orbital migration as a unit in Section \ref{sec5}. The system as a whole could be considered
as a unit,
or it could be considered to be composed of independent subsystems  with  the analysis being  applied to each  separately.
The interaction  between neighbours is supposed to occur through the influence of either one or two retained resonant angles. 
For a system or subsystem in which such  commensurabilities are maintained, the model   enables the
migration times for each planet to be determined once planet masses, circularization times
and the migration time for the innermost planet is specified.
The latter was chosen so that the time scale for the evolution of the system was in the range of $10^{6-7} y.$
as expected for current protoplanetary disks. However, in doing this we required a disk model with mass exceeding that
for a minimum mass solar nebula by a factor $\sim 5.6$ in its inner regions within $1AU.$ 
But note that there is no implication from this concerning the mass distribution at large distances.  

The planet masses for systems such as Kepler 444 are very uncertain as only their radii are known.
To obtain a working model we specified values for the masses such that the migration times for the standard case were
approximately inversely proportional to the planet mass as expected from the simple theory of disk planet interaction.
Migration times and masses determined from these  procedures, along with circularization times
and the current orbital architecture of Kepler 444 were used as input data for detailed numerical simulations.
But it is important to note that results may be scaled so that the initial systems start at an expanded length scale and finish
in configuration resembling the  original one (see Section \ref{Generalscaling}). 

Because of  relatively large deviations from exact  resonance, the migration times found in this way for the standard case
and other cases with the same relative separation from resonances
are such that they would produce  weak convergent  migration  of the system,  contracting  by only  $\sim 1\%$  while 
undergoing significant migration as a whole (see Sections \ref{stan} and \ref{lmsims}).
Such migration rates are also between one and two orders of magnitude smaller than expected from the simplest modelling of disk-planet interaction
 indicating that significant modification along the lines indicated in Section \ref{sec3} is needed.

 This means that the system is unlikely to have reached the 
current architecture from a state where  it's components  were significantly more widely separated.
 In that case these inferred migration rates
could have only applied during the later evolutionary phases. 

Simulations  of the standard case $sacd$ and $s_1ac_1f$  presented in  Sections \ref{stan} and \ref{lmsims} show only a weak coupling between 
planets $2$ and $3$ on account of their relatively large separation from resonance, indicating a tendency of the inner two planets to be decoupled from the outer $3.$  
Although period ratios did not remain strictly constant during the evolution such that Laplace resonances could eventually form, we found that relatively small
adjustments to the migration times  obtained from the analytic model could significantly reduce the variation of the period ratios such that this did not occur. 

In order to study systems showing stronger convergent migration we studied  systems with consecutive pairs, not including planets $3$ and $4,$ an order of magnitude relatively
closer to resonance than in the standard case.  These  can then show significant relative convergence on the time scale for which the system  contracts as a whole.
However, we found again that the inner two planets tended to become detached from the outer $3.$  This  happened
for  migration times determined from the analytic model independently of whether they were obtained assuming the system was split into two separate subsystems
or not (see Sections \ref{allcoupled}  and  \ref{sepsubs}). 

These simulations confirm the view that   if the system underwent significant convergent migration before
reaching the final state the  resonances were closer and the  migration times  shorter during this phase. One could speculate that migration rates slowed and resonances
moved apart as the system approached the inner regions of the disk where a process causing its truncation operated. The latter evolution
is similar to that obtained when  tidal interaction with a central star causes orbital circularization
in the absence of orbital migration (Papaloizou 2011; Lithwick \& Wu 2012; Batygin \& Morbidelli 2013).

Finally we remark that our pilot studies are incomplete as they 
were only carried out in order  to investigate the potential importance of  orbital migration and circularization  
induced by the  tidal interaction  of  a planetary system,  such as Kepler 444,  with a protoplanetary disk modelled in the simplest possible way.
We have only considered processes occurring  when the system is not too far removed from its final configuration and not addressed the issue of  possible
migration from large distances.
Furthermore potentially important processes such as stochastic forces resulting from disk turbulence or planetesimal migration,
which  could occur during the period for which the evolution was studied  and also after the gas disk 
has dispersed,  have been omitted. Nonetheless they indicate that the effects we study could potentially 
 play an important role in producing the current architecture.



\end{document}